\documentclass[11pt]{article}

\usepackage{amsmath}
\usepackage{graphicx}
\usepackage{indentfirst}
\usepackage{amssymb}
\usepackage{cite}
\usepackage{color}
\usepackage{subfigure}
\usepackage{varwidth}
\usepackage{diagbox}
\usepackage[colorlinks=true, linkcolor=red, citecolor=blue, urlcolor=magenta]{hyperref}
\usepackage{xcolor}

\begin{document}

\title{Polarization images of non-topological soliton Bardeen boson stars}

\date{}
\maketitle

\begin{center}
\author{Xiao-Xiong Zeng,}$^{a}$\footnote{E-mail: xxzengphysics@163.com}
\author{Chen-Yu Yang,}$^{b}$\footnote{E-mail: chenyu\_yang2024@163.com}
\author{Ke-Jian He,}$^{b}$\footnote{E-mail: kjhe94@163.com (Corresponding author)}
\author{Li-Fang Li,}$^{c}$\footnote{E-mail:lilifang@imech.ac.cn (Corresponding author) }
\\

\vskip 0.25in
$^{a}$\it{College of Physics and Optoelectronic Engineering, Chongqing Normal University, Chongqing 401331, People's Republic of China}\\
$^{b}$\it{Department of Mechanics, Chongqing Jiaotong University, Chongqing 400000, People's Republic of China}\\
$^{c}$\it{Center for Gravitational Wave Experiment, National Microgravity Laboratory, Institute of Mechanics, Chinese Academy of Sciences, Beijing 100190, People's Republic of China}\\
\end{center}

\vskip 0.6in
{\abstract
{In this study, we investigate the polarized images of non-topological soliton Bardeen boson stars by solving the coupled Einstein nonlinear electrodynamics complex scalar field equations, based on the thin accretion disk model surrounding these compact objects. We focus on the influence of key parameters, including the initial scalar field, magnetic charge, observer inclination angle, and magnetic field configuration, on the resulting polarization characteristics. The results show that the geometry of the magnetic field, particularly the relative strength between the radial \(B_r\) and angular \(B_\theta\) components, plays a crucial role in determining the polarization pattern. Additionally, variations in the scalar field amplitude and magnetic charge significantly affect both the intensity and spatial distribution of the polarization.  These results show that the polarization morphology is sensitive to the spacetime geometry and magnetic field configuration, and provide a qualitative basis for comparing boson stars with black holes.
}}
\thispagestyle{empty}
\newpage
\setcounter{page}{1}\

\section{Introduction}
\label{sec:intro}
In recent years, the development of high-resolution imaging techniques, particularly by the Event Horizon Telescope (EHT) \cite{EventHorizonTelescope:2019dse}, has significantly advanced our understanding of compact astrophysical objects through the acquisition of unprecedented polarized images of black holes, including M87* \cite{EventHorizonTelescope:2021bee,EventHorizonTelescope:2021srq} and Sgr A* \cite{EventHorizonTelescope:2022wkp,EventHorizonTelescope:2022wok}. These observations not only confirm key predictions of general relativity \cite{ALMA:2021axn} but also yield valuable insights into the configurations of magnetic fields and plasma dynamics in extreme gravitational environments. In particular, polarization serves as a critical diagnostic tool, encoding information regarding magnetic field geometry, fluid dynamics, and spacetime curvature\cite{Moscibrodzka:2019adb,Li:2008zr,Moscibrodzka:2017gdx}. In the foundational analytical work presented in \cite{Narayan:2021polarized}, polarization models for Schwarzschild black holes with equatorial accretion disks were established, demonstrating how key parameters such as the observer's inclination angle, magnetic field configuration, and fluid velocity affect the observed polarization signatures. And, this model has subsequently been generalized to Kerr black hole spacetime and to other black hole geometries arising from modified theories of gravity\cite{Gelles:2021kti,Qin:2021xvx,Zhang:2021hit}.
In addition, the Event Horizon Telescope's recent measurements of both linear and circular polarization in Sgr A* \cite{EventHorizonTelescope:2024hpu} have highlighted the significance of investigating alternative gravitational configurations beyond classical black hole models \cite{EventHorizonTelescope:2020qrl}.

Although a substantial body of research has investigated polarized images in black hole spacetimes, including those derived from modified gravity theories and quantum-corrected metrics \cite{
Shi:2024bpm,Yin:2025rao,Delijski:2022jjj,Angelov:2025rut,Chen:2024jkm,Deliyski:2023gik}, alternative compact objects such as boson stars have remained comparatively underexplored in this domain. Boson stars, arising as solutions to Einstein's equations coupled with scalar fields \cite{Kaup:1968zz,Ruffini:1969qy}, offer compelling theoretical models for horizonless compact objects that closely resemble black holes in various phenomenological aspects \cite{vincent2016imaging,rosa2022shadows}. Among these, the non-topological soliton Bardeen boson stars are characterized by self-interacting complex scalar fields and nonlinear electrodynamics \cite{Lee:1986ts,Friedberg:1986tq,Zhao:2025hdg}, providing a robust theoretical framework for exploring quantum gravitational effects and potential observational degeneracies with black holes. These configurations are characterized by static, spherically symmetric metrics that incorporate scalar-field potentials and electromagnetic interactions, both of which influence photon propagation and polarization transport in nontrivial ways. A comprehensive understanding of the polarized radiation emitted by boson stars is essential, as it facilitates their observational differentiation from black holes\cite{Rosa:2025dzq,He:2025qmq,Li:2025awg}.

In diverse black hole backgrounds, relatively well-established analytical and numerical methods have been developed through the consideration of a simple thin accretion disk model\cite{Hou:2022eev,He:2024amh,Zeng:2025tji,Zeng:2025kqw,He:2025hbu,Yang:2026distinguishing}. Based on these works\cite{Zhong:2021mty,Zhang:2023cuw,Zhu:2022amy,Li:2024ctu,yang2025shadow}, our objective is to conduct a study on the polarization images of
non-topological soliton Bardeen boson stars surrounded by thin accretion disks, with the purpose of investigating the disparities between black holes and boson stars in polarization images. We employ a comprehensive theoretical framework in which Einstein's theory of gravity is coupled to nonlinear electrodynamics and a self-interacting complex scalar field \cite{Ayon-Beato:2000mjt}. The scalar potential incorporates mass parameters and interaction coupling strengths that govern the equilibrium configurations of the stellar system \cite{Wang:2023tdz,Zhao:2025hdg}. The spacetime metric is numerically determined under conditions of regularity and asymptotic flatness \cite{Zhao:2025hdg}, which enables the modeling of synchrotron radiation emitted by magnetized plasma in orbital motion.  We generate simulated images via systematic variation of key parameters, such as scalar field amplitude, magnetic charge, observer inclination angle, and magnetic field configuration, to examine the impact of these parameter alterations on the polarization characteristics of non-topological soliton Bardeen boson stars.
The results indicate that the polarization characteristics of boson stars exhibit significant sensitivity to a number of key parameters. In addition, the observer's inclination angle significantly influences the morphology of polarization images , where larger inclination angles give rise to more pronounced asymmetries and increased intensity levels, especially in equatorial regions.
The geometry of the magnetic field particularly the relative strengths of its radial and angular components plays a critical role, giving rise to periodic variations in the electric vector position angle (EVPA) and influencing the spatial distribution of polarization intensity maxima and minima. Furthermore, variations in the initial scalar field amplitude or magnetic charge induce observable shifts in the $Q-U$ plane loops, reflecting corresponding changes in the underlying spacetime structure. Overall, the present analysis provides a qualitative description of how boson star parameters influence polarized image morphology in the dimensionless setup adopted here.

The remainder of this paper is organized as follows. Section~2 introduces the theoretical framework and derives the corresponding field equations. Section~3 describes the numerical methods and boundary conditions. Section~4 presents our results, including the structure of solutions, frozen states and critical horizons, compactness, and light-ring properties. Finally, Section~5 summarizes the main conclusions.

\section{Solutions of  a non-topological soliton Bardeen boson stars}\label{sec2}
We consider a gravitational model that couples Einstein gravity to nonlinear electrodynamics and a self-interacting complex scalar field. The total action governing the system can be written as\cite{Zhao:2025hdg}
\begin{equation}
S = \int d^{4}x\,\sqrt{-g}\,\left(\frac{R}{16\pi} + \mathcal{L}_{\text{EM}} + \mathcal{L}_{\text{SF}}\right),
\label{eq:action}
\end{equation}
where $R$ is the Ricci scalar, $\mathcal{L}_{\text{EM}}$ denotes the nonlinear electromagnetic Lagrangian, and $\mathcal{L}_{\text{SF}}$ represents the complex scalar field contribution.
The electromagnetic Lagrangian is defined as
\begin{equation}
\mathcal{L}_{\text{EM}} = -\frac{3}{2s}
\left[\frac{\sqrt{2q^{2}\mathcal{F}}}{1+\sqrt{2q^{2}\mathcal{F}}}\right]^{\!\frac{5}{2}},
\label{eq:L_EM}
\end{equation}
where $\mathcal{F} \equiv \tfrac{1}{4}F_{\mu\nu}F^{\mu\nu}$ and $F_{\mu\nu} = \partial_{\mu}A_{\nu} - \partial_{\nu}A_{\mu}$. When the magnetic charge $q = 0$ while the complex scalar field $\Phi$ is non-vanishing, the action corresponds to a non-topological soliton black hole. When the scalar field $\Phi$ vanishes but $q \neq 0$, the model reduces to a Bardeen black hole. The parameter $s$ serves to normalize the electromagnetic field strength and determines the ADM mass as $M = q^3 4\pi / (2 s)$. Note that the appearance of $q$ in the Lagrangian is somewhat unusual, making the model specific to particular magnetization states. A non-topological soliton Bardeen boson star discussed below is a stable, localized configuration of a self-interacting complex scalar field coupled to Einstein gravity and a Bardeen-type nonlinear electromagnetic field, whose stability is maintained by the balance between the scalar field's self-interaction and the conserved global U(1) charge, without relying on any topological constraints \cite{Friedberg:1986tq,Lee:1986ts,Bardeen:1968}.
The scalar-field Lagrangian is given by
\begin{equation}
\mathcal{L}_{\text{SF}} = -g^{\mu\nu}\nabla_{\mu}\Phi^{*}\nabla_{\nu}\Phi - U(|\Phi|),
\label{eq:L_SF}
\end{equation}
where the self-interaction potential takes the form
\begin{equation}
U(|\Phi|) = \mu^{2}|\Phi|^{2}\left(1 - 2\eta^{2}|\Phi|^{2}\right)^{2}.
\label{eq:potential}
\end{equation}
Here, $q$ denotes the magnetic charge, $\mu$ is the scalar-field mass parameter, and $\eta$ controls the strength of the scalar self-interaction.
The static, spherically symmetric spacetime is described by the line element
\begin{equation}
ds^{2} = -N(r)\sigma^{2}(r)dt^{2} + \frac{dr^{2}}{N(r)} + r^{2}\left(d\theta^{2} + \sin^{2}\theta\, d\varphi^{2}\right),
\label{eq:metric}
\end{equation}
where $N(r) = 1 - \tfrac{2m(r)}{r}$. By specifying the field configurations, namely, the electromagnetic field and the scalar field \cite{Lee:1986ts,Friedberg:1986tq}, we adopt the following form
\begin{align}
F_{\mu\nu} &= 2q\,\delta^{\theta}_{[\mu}\delta^{\varphi}_{\nu]}\sin\theta, \\
\Phi &= \frac{\phi(r)}{\sqrt{2}}\,e^{-i\omega t},
\label{eq:ansatz}
\end{align}
where $\phi(r)$ is a real-valued function, and $\omega$ represents the field frequency. By varying the action with respect to the metric and scalar field, we obtain the coupled field equations:
\begin{align}
\phi''
&= -\!\left(\frac{2}{r} + \frac{N'}{N} + \frac{\sigma'}{\sigma}\right)\phi'
   - \frac{\phi}{N}\!\left(\frac{\omega^{2}}{N\sigma^{2}} - \frac{\partial U}{\partial \phi^{2}}\right),
\label{eq:phi_eq_alt}\\[6pt]
m'
&= 2\pi r^{2}\!\left(
     \frac{\omega^{2}\phi^{2}}{N\sigma^{2}}
     + N\phi'^{\,2}
     + U(\phi)
     + \mathcal{L}_{\text{EM}}
   \right),
\label{eq:m_eq_alt}\\[6pt]
\frac{\sigma'}{\sigma}
&= 4\pi r\!\left(
     \phi'^{\,2}
     + \frac{\omega^{2}\phi^{2}}{N^{2}\sigma^{2}}
   \right).
\label{eq:sigma_eq_alt}
\end{align}
Boundary conditions ensuring regularity at the origin and asymptotic flatness at spatial infinity are imposed as
\begin{align}
&m(0) = 0,
\qquad \phi'(0) = 0, \qquad \phi(\infty) = 0, \\[4pt]
&\sigma(\infty) = 1,
\qquad m(\infty) = M,
\label{eq:boundary_alt}
\end{align}
For numerical integration, we introduce dimensionless scaled variables through the transformations
\begin{align}
r &\rightarrow \tilde{r}/\mu, \quad
q \rightarrow \tilde{q}/\mu, \quad
\eta \rightarrow \tilde{\eta}\sqrt{4\pi}, \\
\omega &\rightarrow \tilde{\omega}\mu, \quad
s \rightarrow 4\pi\tilde{s}/\mu^{2}, \quad
\phi \rightarrow \tilde{\phi}/\sqrt{4\pi},
\label{eq:scaling}
\end{align}
which render the system scale-invariant under the normalization $\mu \rightarrow 1$ and $4\pi \rightarrow 1$. The radial coordinate is further compactified via the mapping
\begin{equation}
x = \frac{\tilde{r}}{1+\tilde{r}},
\end{equation}
such that $r \in [0,\infty)$ is mapped to $x \in [0,1]$.

We consider a geometrically and optically thin accretion disk located on the equatorial plane of the boson star as the background emission source. Before reaching the observer, photons may interact with the disk multiple times; therefore, the total observed intensity must account for all possible trajectories. Neglecting reflection effects, the total specific flux can be expressed as\cite{Zeng:2026polarization}
\begin{equation}
    F_{\nu_o} = \sum_{n=1}^{N_m} g_n^{3} F_n,
    \label{eq:i}
\end{equation}
where $\nu_o$ denotes the observed frequency, $N_m$ is the maximum number of intersections between the photon trajectories and the accretion disk, and $g_n$ represents the corresponding redshift factor. For the emission profile $F_n$, we adopt the following functional form\cite{Gralla:2020srx}
\begin{equation}
F_n =
\frac{
\exp\!\left[-\tfrac{1}{2}\!\left(\gamma + \operatorname{arcsinh}\!\tfrac{r - v}{\sigma}\right)^{\!2}\right]
}{
\sqrt{(r - v)^{2} + \sigma^{2}}
}.
\end{equation}
Here, $\gamma$ denotes the intensity growth rate, $v$ specifies the location of the emission peak, and $\sigma$ controls the width of the profile. Together, these three parameters ($\gamma$, $v$, and $\sigma$) characterize the thin accretion disk model.
Under the thin-disk assumption, the accreting plasma can be regarded as material moving along geodesics confined to the equatorial plane. The observed polarization thus primarily originates from synchrotron radiation emitted by electrons within the plasma. For an observer comoving with the plasma, the polarization direction of the emitted photons is perpendicular to both the local magnetic field $\vec{B}$ and the photon three-momentum $\vec{k}$. Consequently, the spatial component of the photon polarization vector can be expressed as\cite{Gelles:2021kti,Delijski:2022jjj}
\begin{equation}
\vec{f} = \frac{\vec{k} \times \vec{B}}{|\vec{k}|}.
\end{equation}
In covariant form, the polarization vector of the photon is generally written as
\begin{equation}
f^{\mu} \propto \xi^{\mu\nu\lambda\eta}\,\mu_{\nu}\,k_{\lambda}\,B_{\eta},
\end{equation}
where $\mu_{\nu}$, $k_{\lambda}$, and $B_{\eta}$ denote the four-velocity of the plasma, the photon four-momentum, and the magnetic field, respectively. In this work, we adopt the EHT inspired parameterization for the magnetic field in the fluid frame, expressed as $\vec{B}=(B_r, B_\phi, B_z)$, where $B_r$, $B_\phi$, and $B_z$ represent the radial, azimuthal, and vertical components, respectively. In particular, $\vec{B}=(B_r,0,0)$ and $\vec{B}=(0,B_\phi,0)$ correspond to purely radial and purely azimuthal magnetic fields in the equatorial plane, while $\vec{B}=(0,0,B_z)$ represents a purely vertical magnetic field. Based on the polarization simulation results of M87* by the EHT, we adopt $\vec{B}=(0.87,0.5,0)$ as the fiducial magnetic field configuration for the subsequent analysis \cite{Narayan:2021polarized}. Here the components of $\vec{B}$ are dimensionless quantities normalized in the local fluid frame. They specify only the relative magnetic field geometry rather than an absolute magnetic field strength in physical units. We also consider alternative magnetic field configurations to explore the sensitivity of polarization patterns to variations in the radial, azimuthal, and vertical components. In addition, the polarization vector must satisfy the normalization condition, that is
\begin{equation}
f^{\mu}f_{\mu} = 1.
\end{equation}
In the emitter reference frame, the intensities of linearly polarized and total radiation are denoted by the emissivities $F_P$ and $F_I$, respectively. According to Eq.~(\ref{eq:i}), the total emission can be expressed as
\begin{equation}
F_I = \sum_{n=1}^{N_m} F_n .
\end{equation}
To simplify the calculation, we assume that the radiation intensity depends only on the emission location and is independent of the photon frequency and magnetic field strength. Hence, the emission intensity is purely a function of the radial coordinate $r$, and the linearly polarized emissivity $F_P$ can be rewritten in terms of $F_I$ as
\begin{equation}
F_P = C_0 F_I(r),
\end{equation}
where $0 \leq C_0 \leq 1$ represents the fraction of linear polarization in the total emitted intensity at the emission point. When $C_0 = 1$, the emission is fully linearly polarized. The polarization vector $f^{\mu}$ is parallelly transported along the null geodesic toward the observer, satisfying\cite{Gelles:2021kti}
\begin{equation}
k^{\nu}\nabla_{\nu} f^{\mu} = 0.
\end{equation}
Introducing the affine parameter $\lambda$, this relation can be equivalently expressed as
\begin{equation}
\frac{d f^{\mu}}{d\lambda} + \Gamma^{\mu}_{\nu\lambda} k^{\nu} f^{\lambda} = 0.
\end{equation}
The observed linearly polarized intensity $F_{P\nu_o}$ and the total observed intensity $F_{\nu_o}$ are then given by
\begin{equation}
F_{P\nu_o} = g^{3} F_P = C_0 g^{3} F_I(r), \qquad
F_{\nu_o} = g^{3} F_I,
\end{equation}
where $g$ denotes the redshift factor. To obtain the polarized images of the boson star, it is necessary to specify the observer and construct the corresponding image plane. In this work, we adopt a Zero Angular Momentum observer (ZAMO). Specifically, the ZAMO is located at the spacetime point $(t_o, r_o, \theta_o, \varphi_o)$, within whose local neighborhood an orthonormal tetrad basis is defined as
\begin{equation}
\left(e_{(a)}^{\ \mu}\right) =
\begin{pmatrix}
\sqrt{-\dfrac{1}{g_{tt}}} & 0 & 0 & 0 \\[6pt]
0 & -\sqrt{\dfrac{1}{g_{rr}}} & 0 & 0 \\[6pt]
0 & 0 & \sqrt{\dfrac{1}{g_{\theta\theta}}} & 0 \\[6pt]
0 & 0 & 0 & -\sqrt{\dfrac{1}{g_{\varphi\varphi}}}
\end{pmatrix}.
\end{equation}
The negative signs in the radial and azimuthal components follow the standard ZAMO convention for orthonormal bases and do not influence physical measurements\cite{Hou:2022eev}. In the ZAMO frame, the photon four-momentum is then expressed as
\begin{equation}
p_{(\mu)} = p_{\nu}\, e^{\nu}_{(\mu)},
\end{equation}
where $p_{(\mu)}$ and $p_{\nu}$ denote the photon four-momentum measured in the ZAMO frame and in the Boyer--Lindquist coordinate system, respectively. At the observer location, we further define a celestial coordinate system $(X, Y)$, which can be mapped onto the Cartesian coordinates $(x, y)$ of the image plane. The relation between the photon four-momentum $p_{(\mu)}$ and the celestial coordinates $(X, Y)$ is given by\cite{Hu:2020usx}
\begin{equation}
\cos X = \frac{p^{(r)}}{p^{(t)}}, \qquad
\tan Y = \frac{p^{(\varphi)}}{p^{(\theta)}}.
\end{equation}
On the image plane $(x, y)$, the Cartesian coordinates are related to the celestial coordinates through
\begin{equation}
x = -2\tan\!\frac{X}{2}\,\sin Y, \qquad
y = -2\tan\!\frac{X}{2}\,\cos Y.
\end{equation}
This work focuses on two observable quantities in polarized imaging, i.e., the direction of the polarization vector on the observer celestial sphere and the polarization intensity. On the image plane, the projection of the polarization vector $f$ satisfies
\begin{equation}
f^{(x)} = f^{\mu} \cdot e_{x} = -f^{\mu} \cdot e_{\varphi}, \qquad
f^{(y)} = f^{\mu} \cdot e_{y} = -f^{\mu} \cdot e_{\theta}.
\end{equation}
Because $\vec{f}$ and $-\vec{f}$ describe the same state of linear polarization, the direction of the polarization vector is standardized by requiring $f^{(y)} > 0$, and the electric vector position angle (EVPA), denoted by $\Phi_{E}$, is restricted to the range $\Phi_{E} \in (0, \pi)$. The total observed linear polarization intensity is obtained by summing the polarized emission from all emission points on the equatorial plane. According to the definitions of the Stokes parameters $Q$ and $U$, the total linear polarization components are given by\cite{Huang:2024bar}
\begin{align}
Q_{\mathrm{all}} &= \sum_{n=1}^{N_m} Q_n = \sum_{n=1}^{N_m} g_n^{3} F_{Pn}
\left[\left(f^{(x)}_n\right)^{2} - \left(f^{(y)}_n\right)^{2}\right],\\
U_{\mathrm{all}} &= \sum_{n=1}^{N_m} U_n = \sum_{n=1}^{N_m} g_n^{3} F_{Pn}
\left(2 f^{(x)}_n f^{(y)}_n\right).
\end{align}
Consequently, the total observed polarization intensity and EVPA can be written as
\begin{equation}
F_{P\nu_o} = \sqrt{Q_{\mathrm{all}}^{2} + U_{\mathrm{all}}^{2}}, \qquad
\Phi_{E} = \frac{1}{2}\arctan\!\frac{U_{\mathrm{all}}}{Q_{\mathrm{all}}}.
\end{equation}
Based on the above formulation, the polarized images of boson stars can be numerically constructed on the image plane. In real numerical calculations, we adopt a resolution of $300 \times 300$ pixels, and integrate the affine parameter over the range $0$--$3000$. We use an adaptive high-order integration method, with strict local error control, ensuring good overall convergence. For details of the numerical implementation, see Ref.~\cite{yang2026shadow}. Additionally, all calculations in this work are performed in natural units, and the parameters used in the images are dimensionless. The present analysis is intended to compare the normalized image morphology, polarization intensity distribution, and EVPA patterns within the same thin accretion disk model. Since no physical scale for the magnetic field, plasma density, electron temperature, or radiative transfer normalization is specified, the dimensionless magnetic field components should not be interpreted as absolute magnetic field strengths in physical units. A direct comparison with specific sources such as Sgr A* or M87* would require an additional physical calibration, which is beyond the scope of the present work.

\section{Polarization characteristics of boson stars}\label{sec4}
Building upon the theoretical framework of polarized synchrotron emission in curved spacetime, we now investigate the characteristic polarization patterns of boson stars, as illustrated in Figure~1. These images display the polarization structures produced under the thin accretion disk model with a field of view of $\gamma_{\mathrm{fov}} = 3.5^\circ$. The results are arranged in a $4 \times 4$ matrix, where the rows correspond to boson star configurations $S_{\phi_0}\mathrm{BS1}$ through $S_{\phi_0}\mathrm{BS4}$ with initial scalar field amplitudes $\phi_0 = 0.65$, $0.70$, $0.75$, and $0.80$ (from top to bottom). Each row presents polarization maps for observation angles $\theta = 1^\circ$, $30^\circ$, $60^\circ$, and $80^\circ$ (from left to right), with the parameters fixed at $s = 0.6$ and $q = 0.5$.

\begin{figure}[!h]
\centering 
\subfigure[$\phi_0=0.65,\theta=1^{\circ}$]{\includegraphics[scale=0.375]{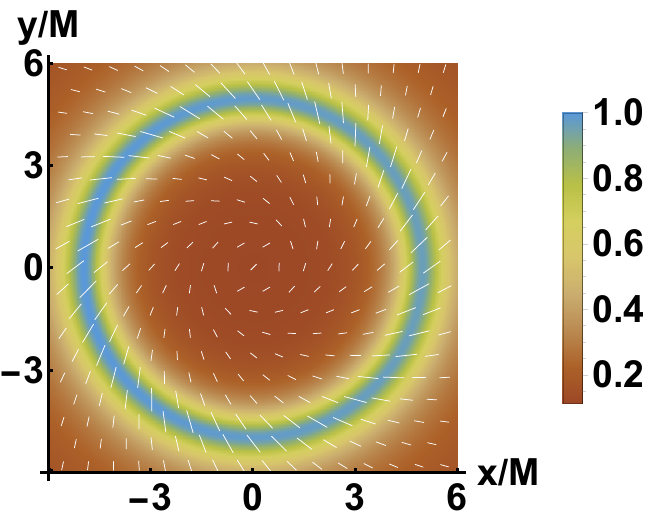}}
\subfigure[$\phi_0=0.65,\theta=30^{\circ}$]{\includegraphics[scale=0.375]{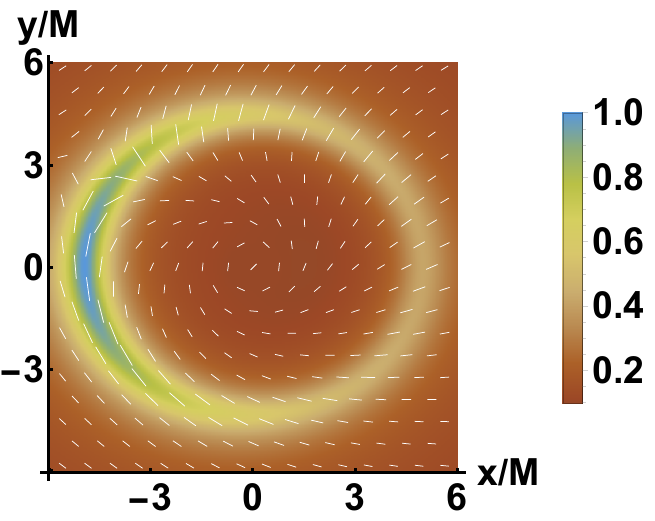}}
\subfigure[$\phi_0=0.65,\theta=60^{\circ}$]{\includegraphics[scale=0.375]{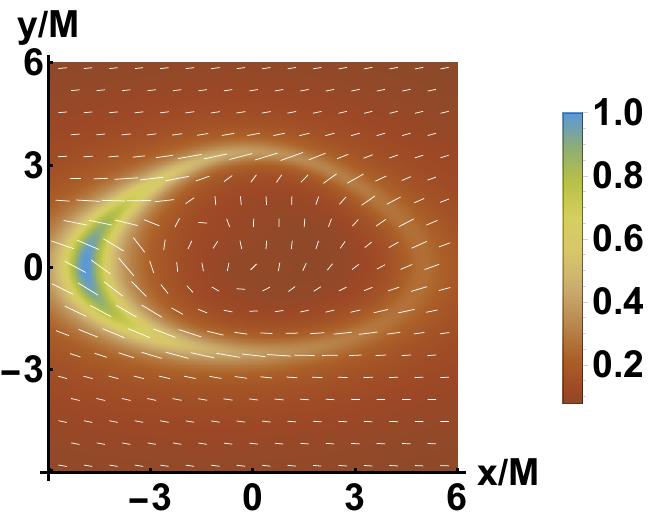}}
\subfigure[$\phi_0=0.65,\theta=80^{\circ}$]{\includegraphics[scale=0.375]{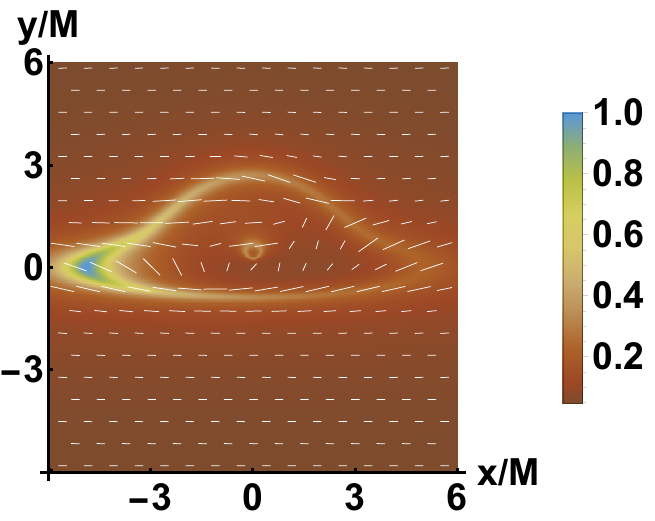}}
\subfigure[$\phi_0=0.7,\theta=1^{\circ}$]{\includegraphics[scale=0.375]{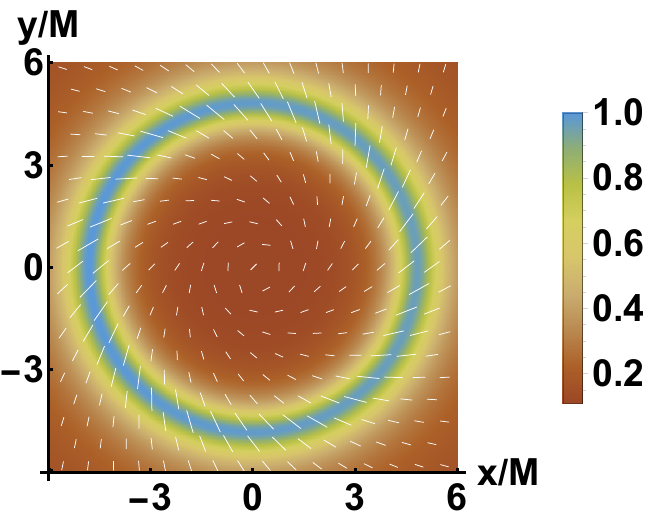}}
\subfigure[$\phi_0=0.7,\theta=30^{\circ}$]{\includegraphics[scale=0.375]{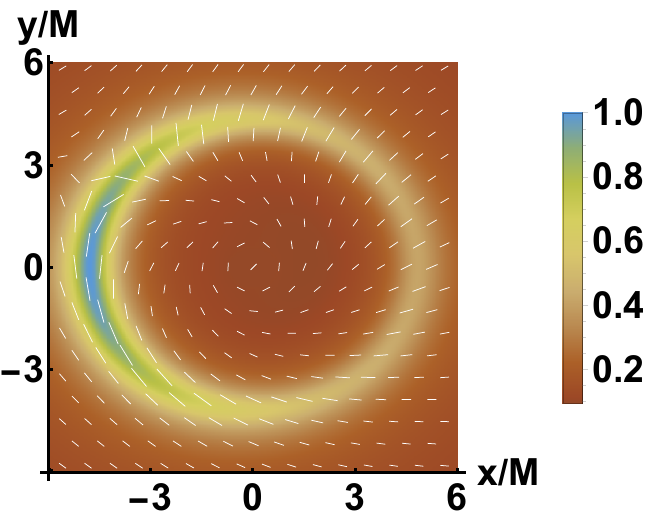}}
\subfigure[$\phi_0=0.7,\theta=60^{\circ}$]{\includegraphics[scale=0.375]{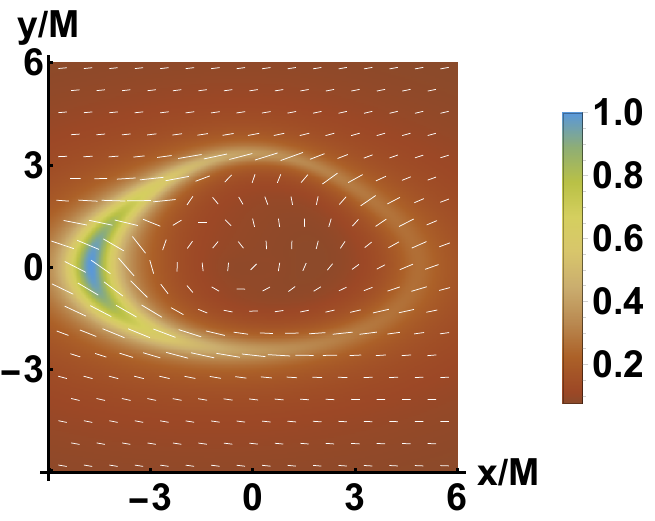}}
\subfigure[$\phi_0=0.7,\theta=80^{\circ}$]{\includegraphics[scale=0.375]{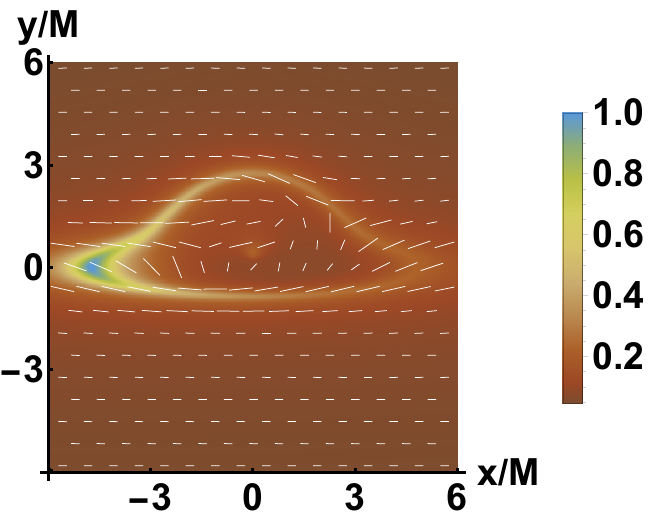}}
\subfigure[$\phi_0=0.75,\theta=1^{\circ}$]{\includegraphics[scale=0.375]{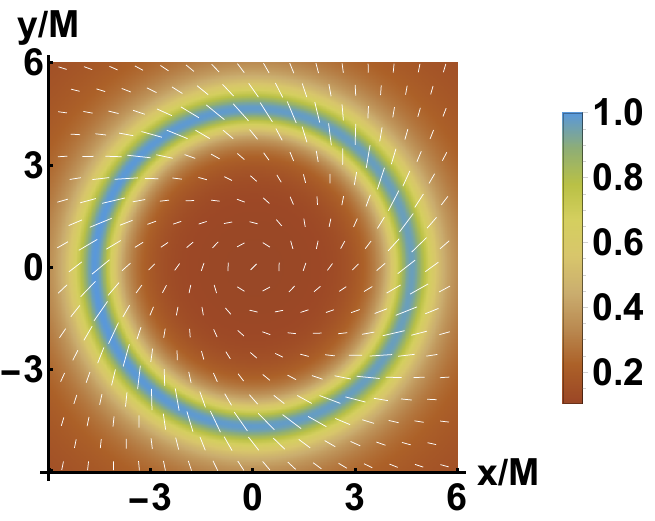}}
\subfigure[$\phi_0=0.75,\theta=30^{\circ}$]{\includegraphics[scale=0.375]{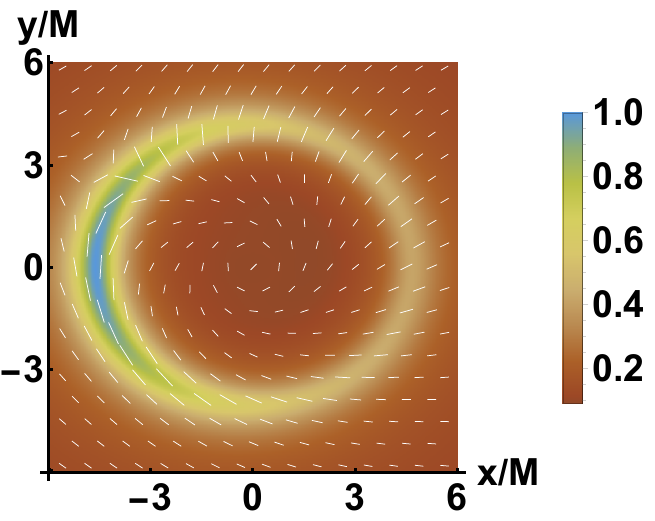}}
\subfigure[$\phi_0=0.75,\theta=60^{\circ}$]{\includegraphics[scale=0.375]{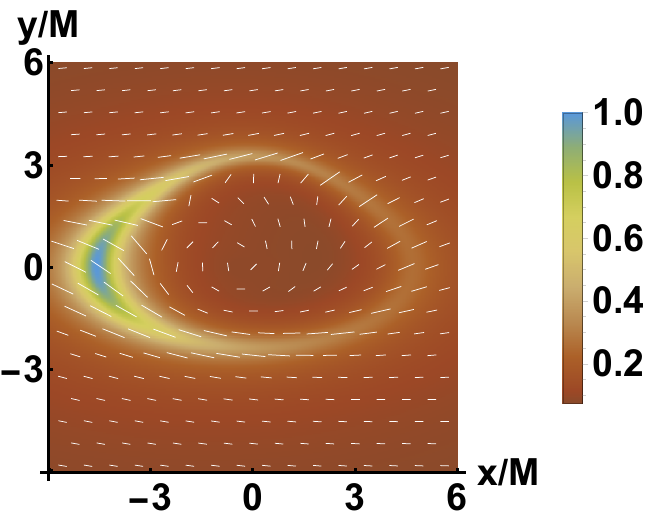}}
\subfigure[$\phi_0=0.75,\theta=80^{\circ}$]{\includegraphics[scale=0.375]{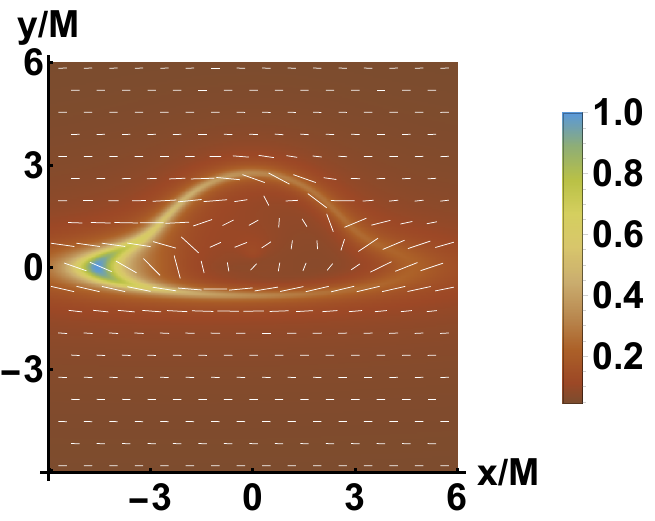}}
\subfigure[$\phi_0=0.8,\theta=1^{\circ}$]{\includegraphics[scale=0.375]{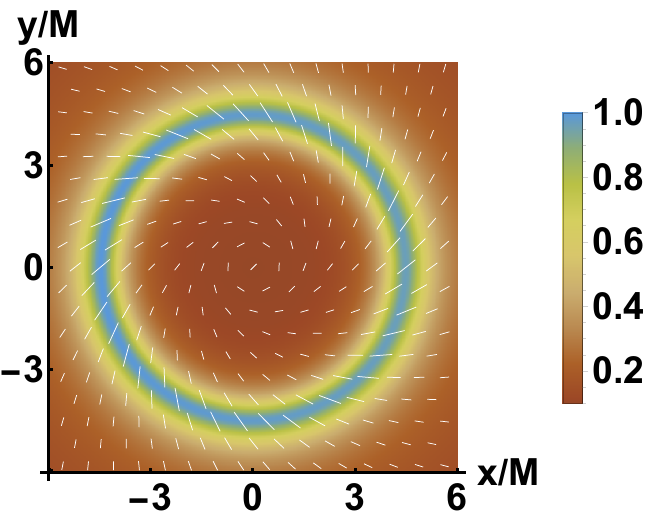}}
\subfigure[$\phi_0=0.8,\theta=30^{\circ}$]{\includegraphics[scale=0.375]{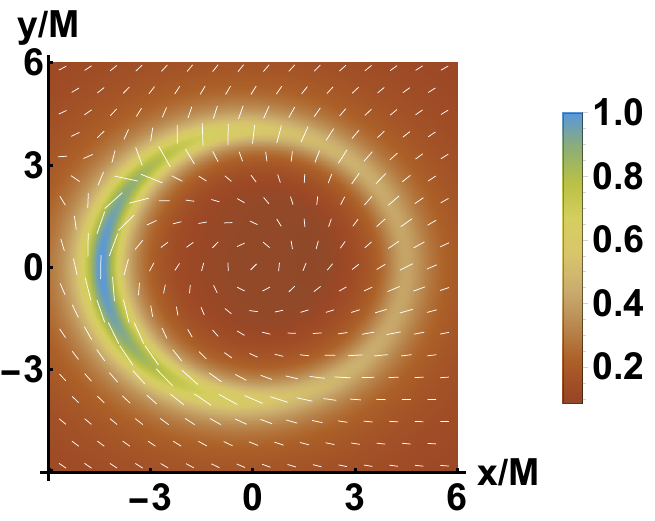}}
\subfigure[$\phi_0=0.8,\theta=60^{\circ}$]{\includegraphics[scale=0.375]{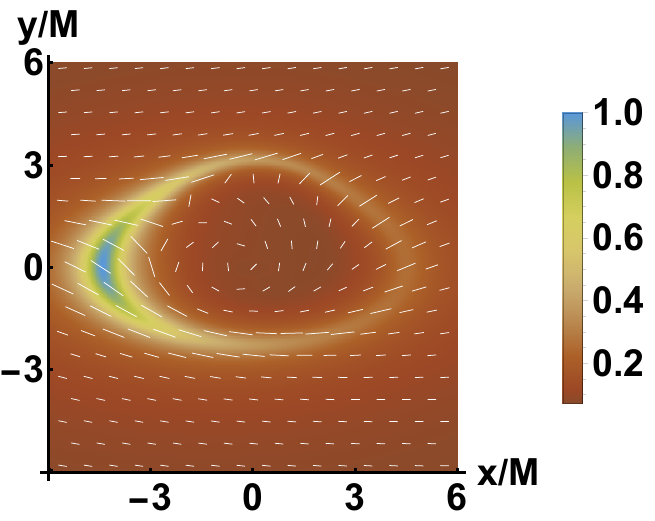}}
\subfigure[$\phi_0=0.8,\theta=80^{\circ}$]{\includegraphics[scale=0.375]{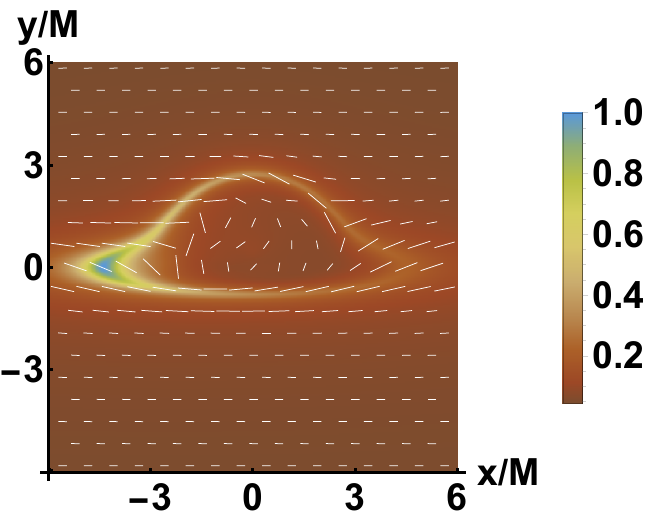}}
\caption{\label{fig1} The polarization images of boson stars under a thin accretion disk model with a field of view angle $\gamma_{_{fov}}=3.5^\circ$. The images are organized into four rows corresponding to boson stars $S_{\phi_0}BS1$, $S_{\phi_0}BS2$, $S_{\phi_0}BS3$, $S_{\phi_0}BS4$ with initial scalar field values $\phi_0=0.65,0.7,0.75,0.8$
(top to bottom). Each row shows four panels with observation inclination angles $\theta=1^\circ, 30^\circ, 60^\circ, 80^\circ$(left to right).  }
\end{figure}

Figure~1 presents the polarization images obtained under the thin accretion disk model with a field of view of $\gamma_{\mathrm{fov}} = 3.5^\circ$, systematically arranged to illustrate how the scalar field strength and the observation angle affect the polarization characteristics. The polarization structure demonstrates a significant dependence on the viewing inclination. Specifically, near face on orientations ($\theta = 1^\circ$) display approximately rotationally symmetric polarization vectors that form concentric patterns around the central object.
As the inclination increases to $\theta = 80^\circ$, the polarization pattern becomes markedly asymmetric with a pronounced directional preference, and the maximum polarization intensity appears along the projected equatorial plane. Furthermore, boson stars with stronger initial scalar field amplitudes ($\phi_0 = 0.75$ and $0.80$) exhibit expanded polarization regions compared with weaker-field configurations ($\phi_0 = 0.65$ and $0.70$). In addition, the electric vector position angle demonstrates a systematic rotation with azimuthal angle, forming spiral-like structures that are most prominent at intermediate inclinations and for higher scalar field amplitudes $\phi_0$. The maximum polarization intensity consistently occurs in the southwestern quadrant at large viewing angles. These results indicate that boson stars share qualitative polarization features with quantum-corrected black holes, such as the expansion of polarized regions and the reduction of intensity with increasing compactness parameters, while the distinctive EVPA spirals provide qualitative indicators for comparing boson stars with conventional compact objects~\cite{Guo:2024bzq}.

\begin{figure}[!h]
\centering 
\subfigure[$\theta=20^{\circ}$]{\includegraphics[scale=0.475]{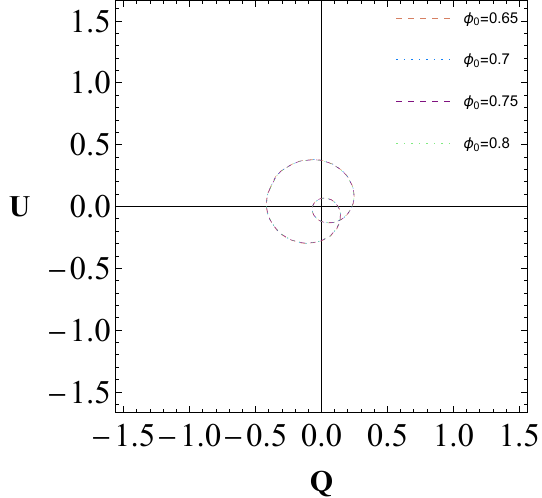}}
\subfigure[$\theta=30^{\circ}$]{\includegraphics[scale=0.475]{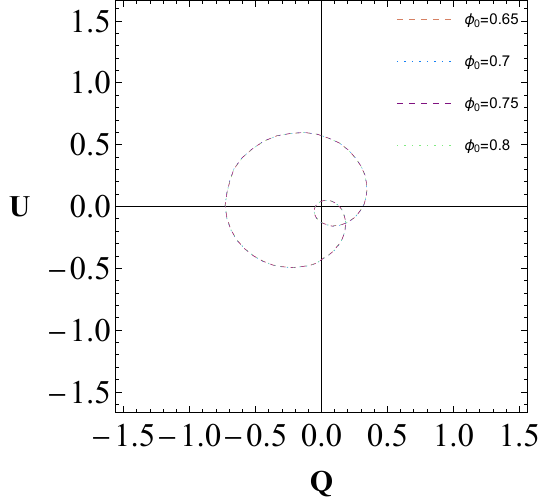}}
\subfigure[$\theta=40^{\circ}$]{\includegraphics[scale=0.475]{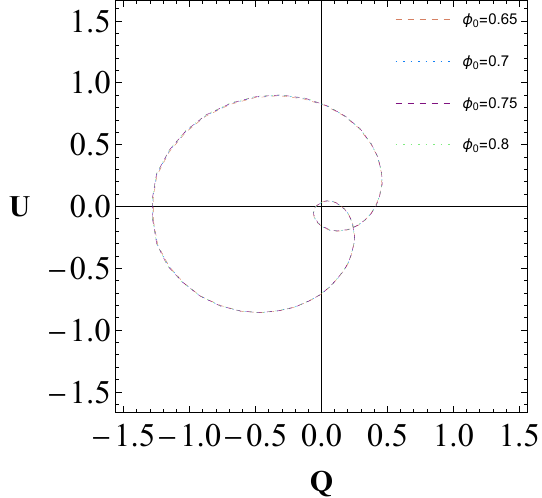}}
\subfigure[$B=(0,0.87,0.1,0)$]{\includegraphics[scale=0.475]{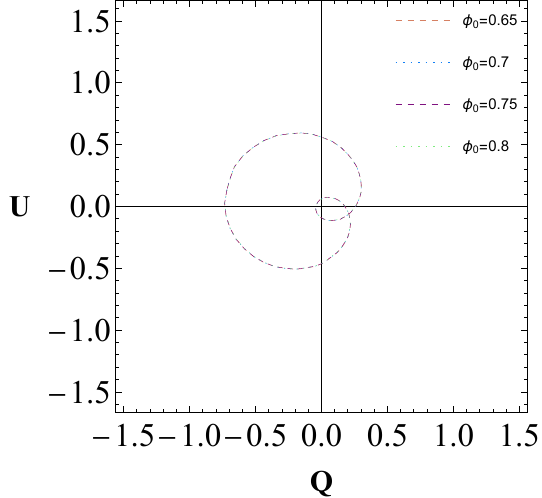}}
\subfigure[$B=(0,0.87,0.3,0)$]{\includegraphics[scale=0.475]{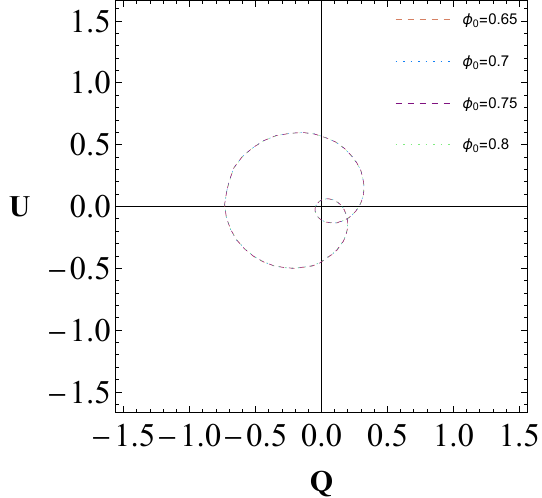}}
\subfigure[$B=(0,0.87,0.5,0)$]{\includegraphics[scale=0.475]{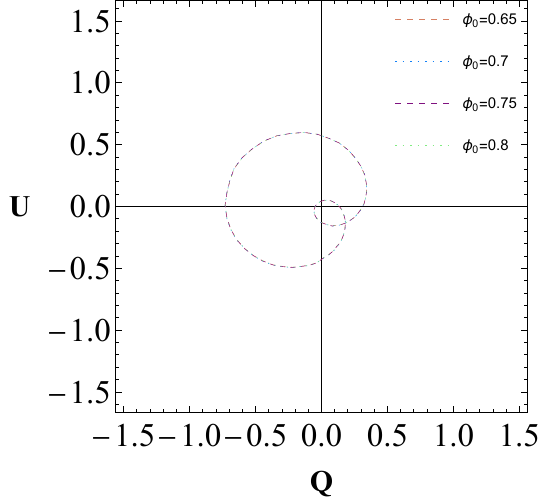}}
\subfigure[$B=(0,0.2,0.5,0)$]{\includegraphics[scale=0.475]{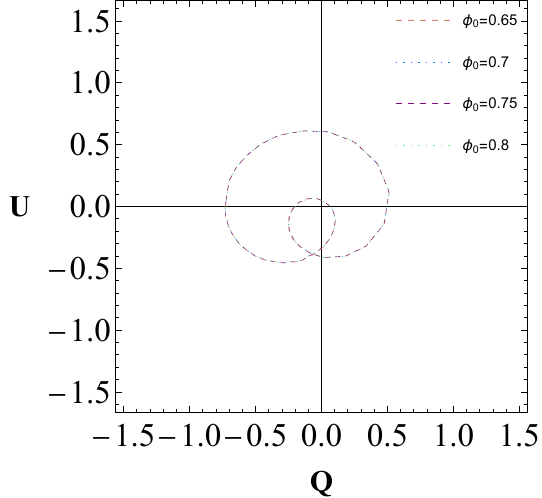}}
\subfigure[$B=(0,0.5,0.5,0)$]{\includegraphics[scale=0.475]{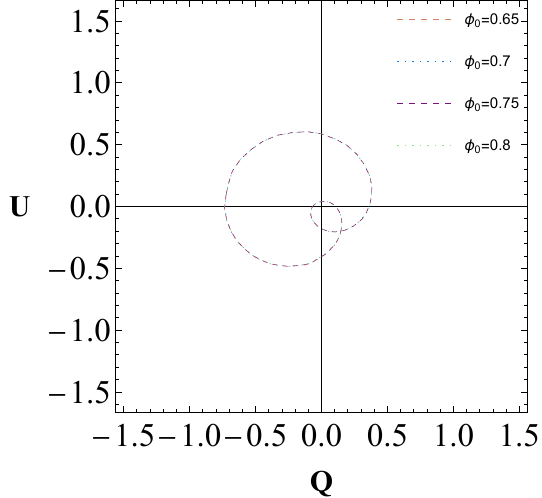}}
\subfigure[$B=(0,0.8,0.5,0)$]{\includegraphics[scale=0.475]{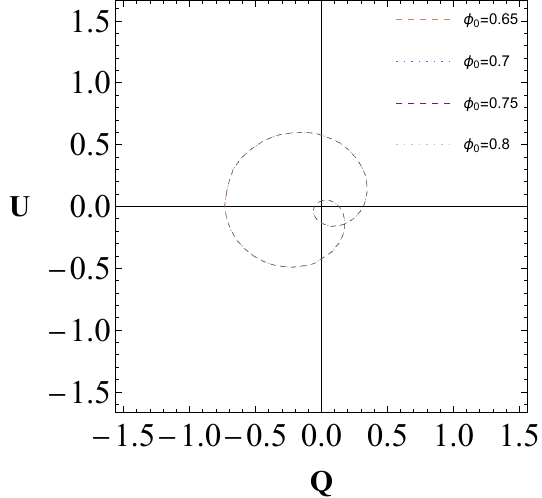}}
\caption{\label{fig2}  The $Q-U$ planes for boson stars \( S_{\phi_0}\mathrm{BS1} \) to \( S_{\phi_0}\mathrm{BS4} \) under varying conditions. The top row varies observation angles \(\theta = 20^\circ\), \(30^\circ\), \(40^\circ\) (left to right) with fixed magnetic field \( B = (0, 0.87, 0.5, 0) \). The middle row varies the \( B_\theta \) component as \( B_\theta = 0.1\), \(0.3\), \(0.5\) (left to right) at fixed \(\theta = 30^\circ\). The bottom row varies the \( B_r \) component as \( B_r = 0.2\), \(0.5\), \(0.8\) (left to right) at fixed \(\theta = 30^\circ\). }
\end{figure}
Next, we examine the distributions of the Stokes parameters in the $Q$-$U$ plane for the same boson star configurations under varying physical conditions, as illustrated in Figure~\ref{fig2}. The $Q$-$U$ plane provides a complete description of the polarization state of the radiation: the radial distance from the origin represents the polarized intensity, while the azimuthal angle encodes the electric vector position angle.
The polarization characteristics of boson stars, quantified in the $Q$-$U$ plane, exhibit three fundamental dependencies. Increasing the observational inclination angle from $\theta = 20^\circ$ to $40^\circ$ produces progressive elliptical deformation and azimuthal rotation of the $Q$-$U$ loops. While near-circular symmetry persists at $\theta = 20^\circ$, the loops become elongated and shift preferentially toward the $Q < 0$ quadrant at $\theta = 40^\circ$, indicating a systematic realignment of polarization vectors toward the southeastern region of the image plane-an effect amplified by the extended scalar cores of the boson stars. Furthermore, strengthening the vertical poloidal magnetic component $B_\theta$ from 0.1 to 0.5 induces loop contraction with reduced eccentricity, yielding compact quasi-circular distributions and a migration of loop centers toward $U > 0$, reflecting a magnetic-field-driven realignment of the electric vector position angles perpendicular to the accretion-disk normal. Conversely, enhancing the radial magnetic component $B_r$ from 0.2 to 0.8 results in loop shearing and asymmetric distortions, culminating in fragmented arcs within the $Q > 0$ half-plane at $B_r = 0.8$. This loss of polarization coherence becomes more pronounced for high-scalar-amplitude configurations ($S_{\phi_0}\mathrm{BS4}$), accompanied by a substantial reduction in the maximum polarized intensity.

These trends contrast with those observed in quantum-corrected black holes\cite{Guo:2024bzq}, where the $Q$-$U$ loops contract uniformly without fragmentation, emphasizing the critical role of horizonless topology in the depolarization mechanisms of boson stars. In contrast to quantum-deformed Schwarzschild black holes, in which the $Q$-$U$ loops shrink uniformly under metric deformation without topological disruption, boson stars exhibit field-dependent loop fragmentation and polarization decoherence. This distinction originates from the absence of an event horizon, which allows magnetic fields to penetrate the scalar core and interfere with polarization transport. The $Q$-$U$ plane therefore provides a useful qualitative indicator: fragmented loops are associated with horizonless compact objects in the present model, whereas uniformly contracted but coherent loops characterize quantum-corrected black holes. These results demonstrate that boson star polarization is highly sensitive to both the observer's geometry and the magnetic-field topology, with strong radial fields producing irreversible depolarization at high scalar amplitudes.

\section{The influence of other parameters on polarization characteristics}
In this section, we will focus on discussing the influence of variations in other related parameters on the polarization characteristics of the non-topological soliton Bardeen boson stars. Different parameter configurations can influence the structure of boson stars, thereby inducing corresponding alterations in the geometry of spacetime. It is essential to extract such underlying information from the polarization characteristics of boson stars.

\begin{figure}[!h]
\centering 
\subfigure[$s=0.2,\theta=1^{\circ}$]{\includegraphics[scale=0.375]{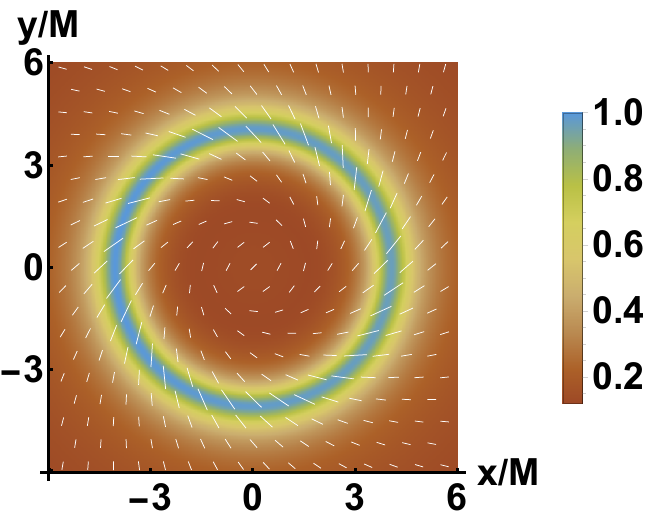}}
\subfigure[$s=0.2,\theta=30^{\circ}$]{\includegraphics[scale=0.375]{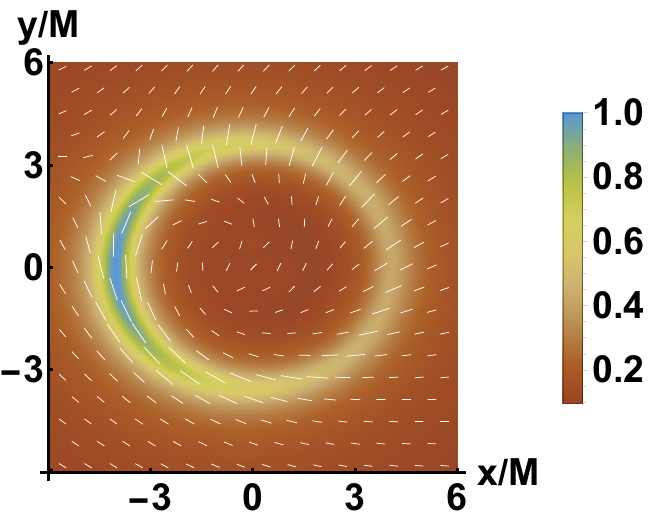}}
\subfigure[$s=0.2,\theta=60^{\circ}$]{\includegraphics[scale=0.375]{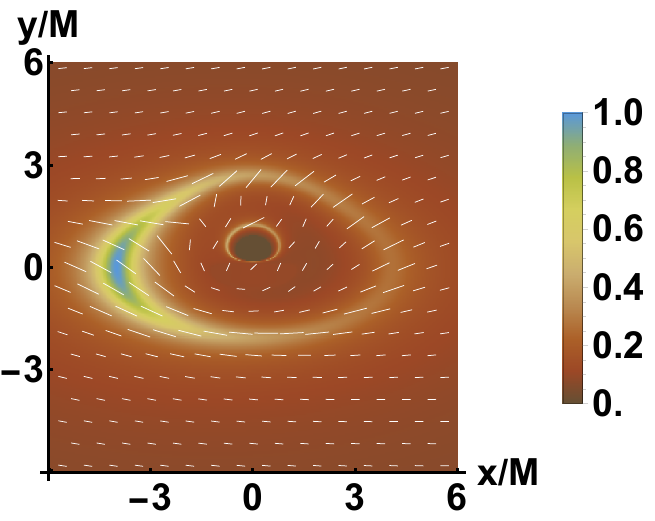}}
\subfigure[$s=0.2,\theta=80^{\circ}$]{\includegraphics[scale=0.375]{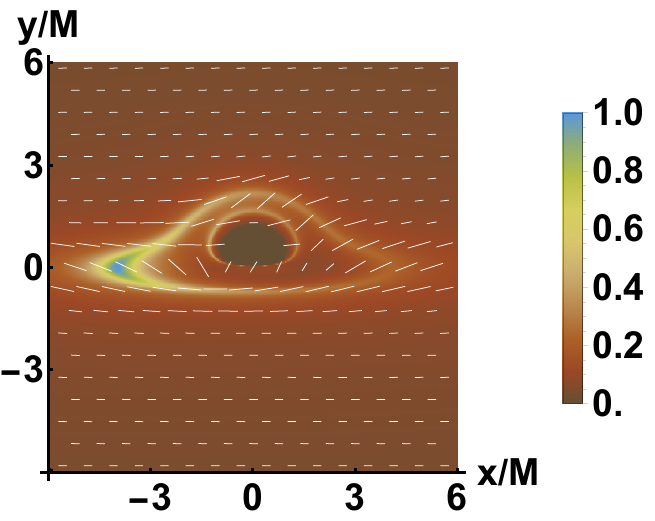}}
\subfigure[$s=0.21,\theta=1^{\circ}$]{\includegraphics[scale=0.375]{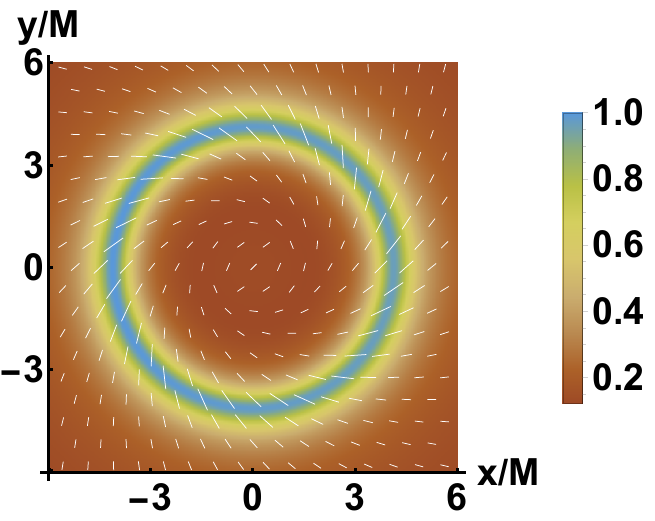}}
\subfigure[$s=0.21,\theta=30^{\circ}$]{\includegraphics[scale=0.375]{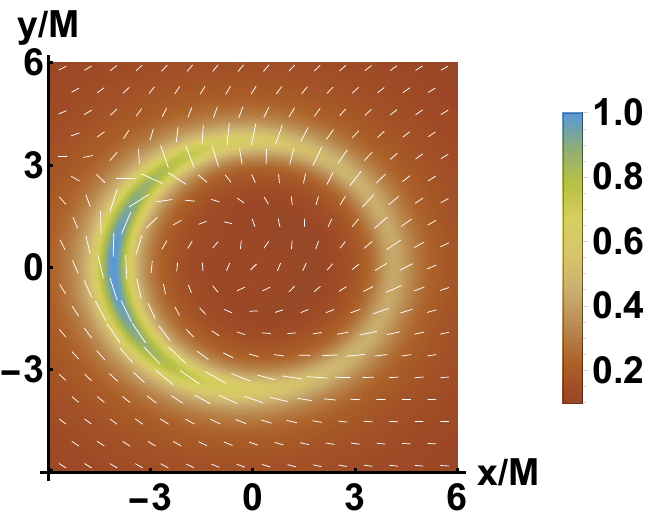}}
\subfigure[$s=0.21,\theta=60^{\circ}$]{\includegraphics[scale=0.375]{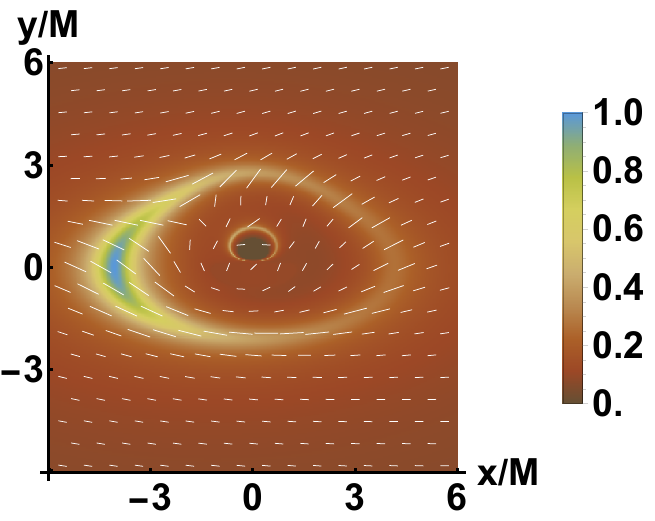}}
\subfigure[$s=0.21,\theta=80^{\circ}$]{\includegraphics[scale=0.375]{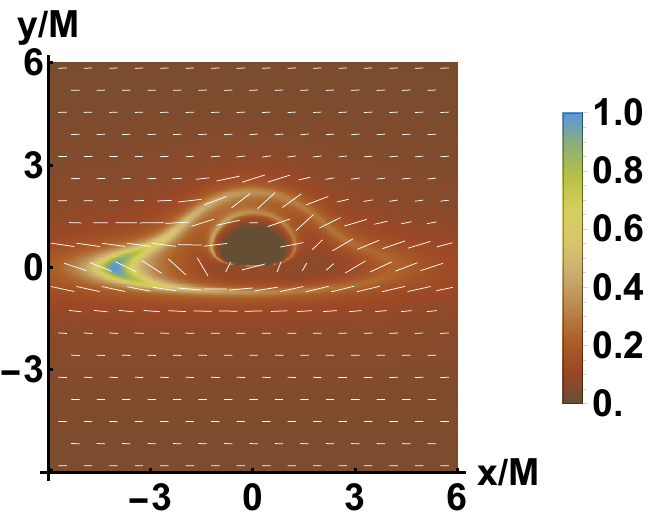}}
\subfigure[$s=0.22,\theta=1^{\circ}$]{\includegraphics[scale=0.375]{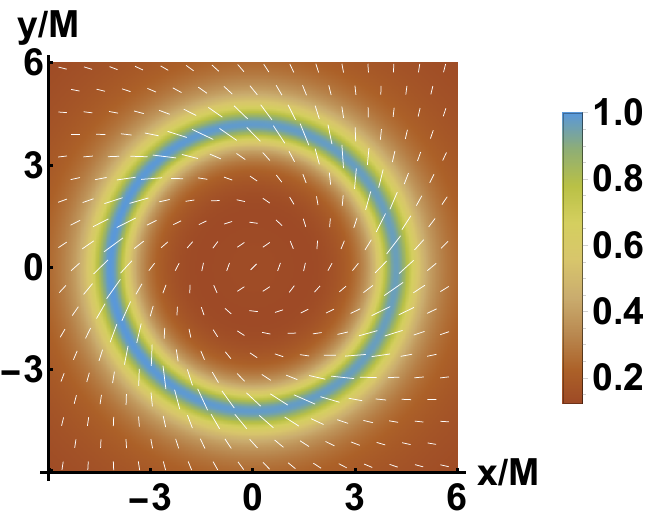}}
\subfigure[$s=0.22,\theta=30^{\circ}$]{\includegraphics[scale=0.375]{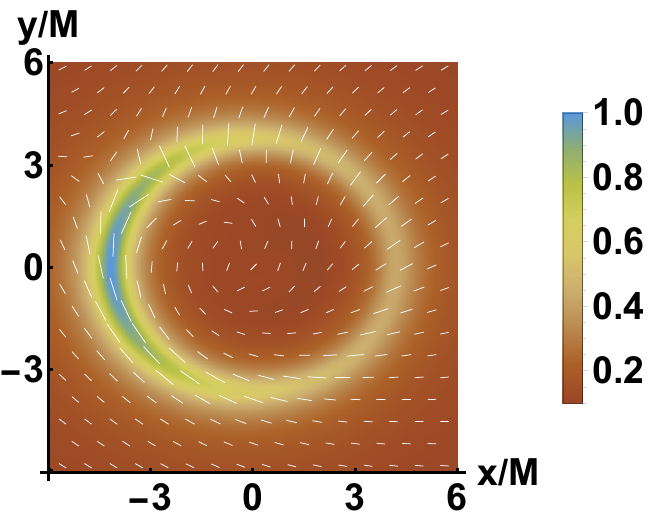}}
\subfigure[$s=0.22,\theta=60^{\circ}$]{\includegraphics[scale=0.375]{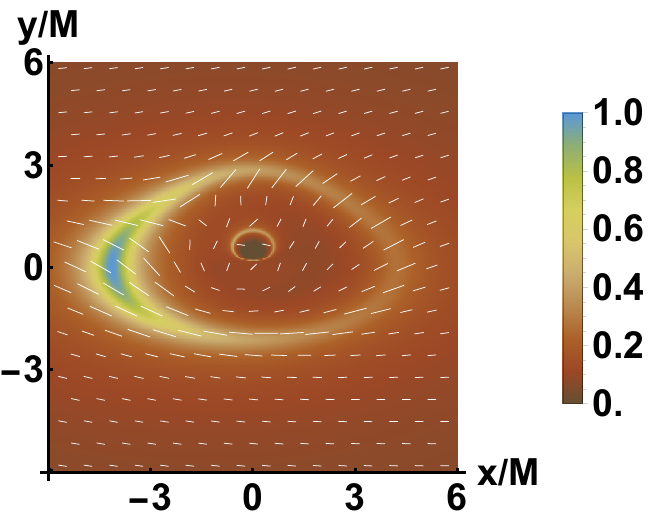}}
\subfigure[$s=0.22,\theta=80^{\circ}$]{\includegraphics[scale=0.375]{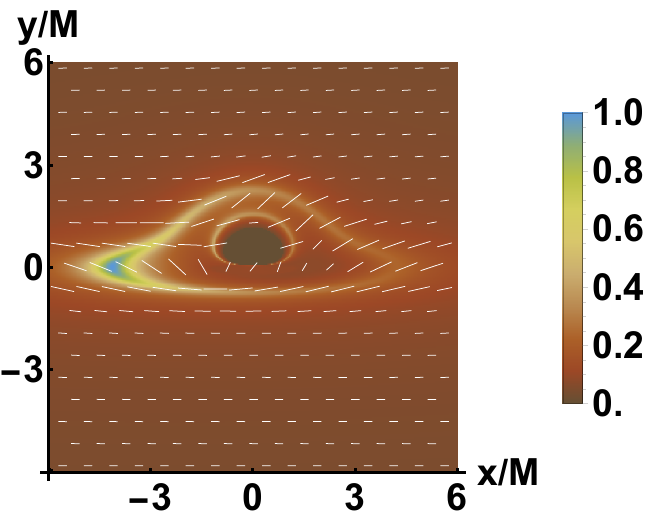}}
\subfigure[$s=0.25,\theta=1^{\circ}$]{\includegraphics[scale=0.375]{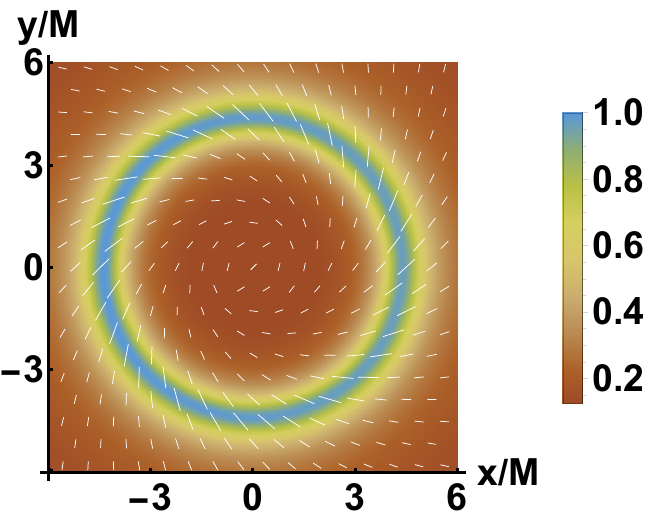}}
\subfigure[$s=0.25,\theta=30^{\circ}$]{\includegraphics[scale=0.375]{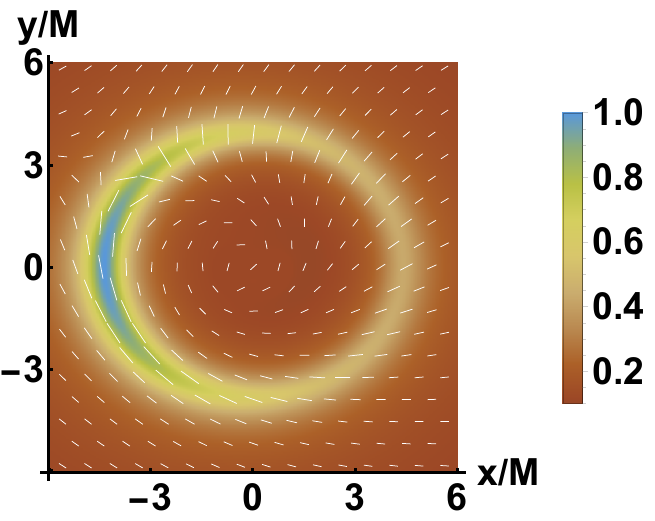}}
\subfigure[$s=0.25,\theta=60^{\circ}$]{\includegraphics[scale=0.375]{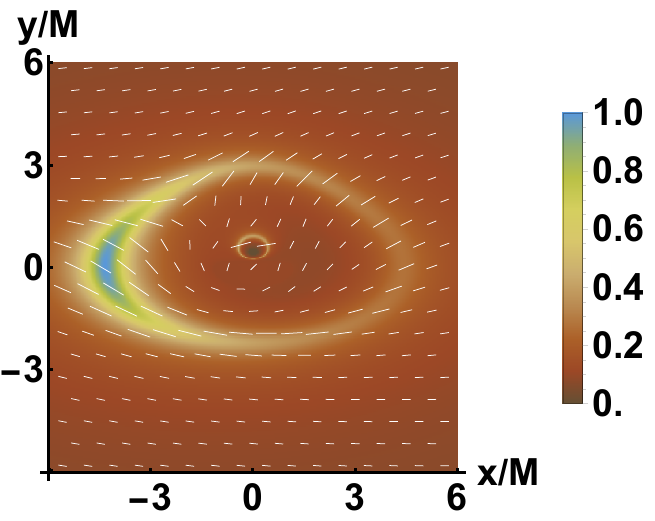}}
\subfigure[$s=0.25,\theta=80^{\circ}$]{\includegraphics[scale=0.375]{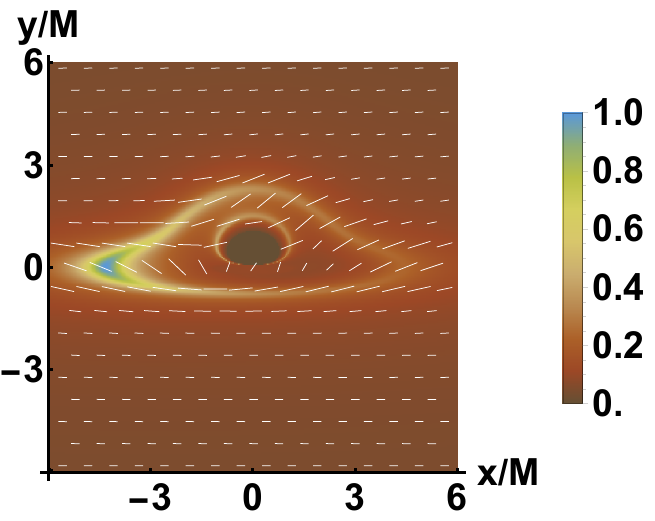}}
\caption{\label{fig3} The optical appearance (non-polarimetric) of boson stars under a thin accretion disk model with field of view \(\gamma_{\mathrm{fov}} = 3.5^\circ\). Images are divided into four rows for boson stars \(S_s\mathrm{BS1}\), \(S_s\mathrm{BS2}\), \(S_s\mathrm{BS3}\), and \(S_s\mathrm{BS4}\) with free parameter values \(s = 0.2\), \(0.21\), \(0.22\), \(0.25\) (top to bottom). Each row shows \(\theta = 1^\circ\), \(30^\circ\), \(60^\circ\), \(80^\circ\) (left to right). }
\end{figure}

Figure~\ref{fig3} illustrates the influence of the free parameters $s$ and $\theta$ on the polarization image of the Boson star. The background of the image is the optical image of the Boson star under a thin accretion disk, with the Boson star located at the center of the image. The colorbar reflects the variation in light intensity, with blue indicating the maximum intensity and brown indicating the minimum intensity. When $\theta = 1^\circ$ (first column), a circular bright ring appears in the image, corresponding to the direct image. At this point, since the observer's line of sight is perpendicular to the equatorial plane, gravitational redshift dominates. As $s$ increases, the shape of the direct image remains unchanged, but the overall size increases. When $\theta$ increases to $30^\circ$ (second column), the direct image becomes asymmetric due to the enhanced Doppler redshift caused by relative motion, with the intensity on the left side greater than that on the right. As $\theta$ increases further (third and fourth columns), the direct image gradually evolves into a cap-like shape. The above phenomena suggest that the observer's inclination angle $\theta$ affects the overall shape of the optical image, while $s$ alters the size of the star's optical image.

The white line segments in the figure represent the polarization vectors, with their length and direction respectively reflecting the total linear polarization intensity and the EVPA. It can be observed from the figure that the polarization intensity in the brighter regions is significantly enhanced compared to the darker regions. For example, in Figure~\ref{fig3}(b), the Doppler effect causes the light intensity on the left side of the direct image to be greater than that on the right, thus displaying a distinct polarization intensity distribution on the left side. It is worth noting that, for black holes, since radiation cannot escape beyond the event horizon, polarization effects are theoretically unobservable in its interior region. However, the central region of the Boson star may exhibit strong polarization effects (first and second columns).

Figure~4 presents the $Q-U$ plane polarization characteristics of boson stars with varying nonlinear electrodynamics parameters $s$ ($S_s\mathrm{BS1}$-$S_s\mathrm{BS4}$, corresponding to $s = \{0.20, 0.21, 0.22, 0.25\}$), systematically examining their response to different magnetic-field configurations and observational geometries. The figure consists of three rows, each demonstrating a key parametric dependence. The first row  illustrates the effect of inclination angle at a fixed magnetic field $B = (0,\,0.87,\,0.5,\,0)$, showing progressive loop deformation as $\theta$ increases from $20^\circ$ to $40^\circ$. This manifests as elliptical distortion and azimuthal rotation toward the $Q < 0$ quadrant, with the deformation amplitude positively correlated with the magnitude of $s$. Configurations with higher $s$ values ($S_s\mathrm{BS3}$, $S_s\mathrm{BS4}$) exhibit enhanced sensitivity to inclination changes, developing pronounced asymmetries at $\theta = 40^\circ$. The second row illustrates the effect of poloidal-field strengthening, showing systematic loop contraction as the vertical component $B_\theta$ increases from 0.1 to 0.5 at fixed $\theta = 30^\circ$. This compression reduces loop eccentricity while driving the loop centers to migrate toward $U > 0$. Notably, the contraction rate depends nonlinearly on $s$: configurations with higher $s$ values ($S_s\mathrm{BS4}$) exhibit significantly amplified contraction compared with lower-$s$ counterparts ($S_s\mathrm{BS1}$), indicating an $s$-modulated sensitivity to poloidal magnetic fields. The third row  depicts the effect of radial-field enhancement, revealing pronounced topological fragmentation as $B_r$ increases from 0.2 to 0.8 at $\theta = 30^\circ$. While all configurations develop asymmetric distortions, high $s$ boson stars ($S_s\mathrm{BS3}$, $S_s\mathrm{BS4}$) undergo complete loop disintegration into sheared arcs confined to the $Q > 0$ half-plane at $B_r = 0.8$. This fragmentation coincides with substantial depolarization, as the maximum polarized intensity decreases by approximately 50\% relative to the low-$B_r$ configurations.

\begin{figure}[!h]
\centering 
\subfigure[$\theta=20^{\circ}$]{\includegraphics[scale=0.475]{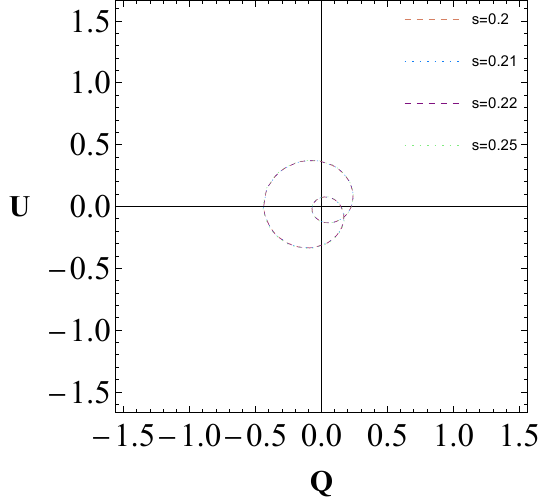}}
\subfigure[$\theta=30^{\circ}$]{\includegraphics[scale=0.475]{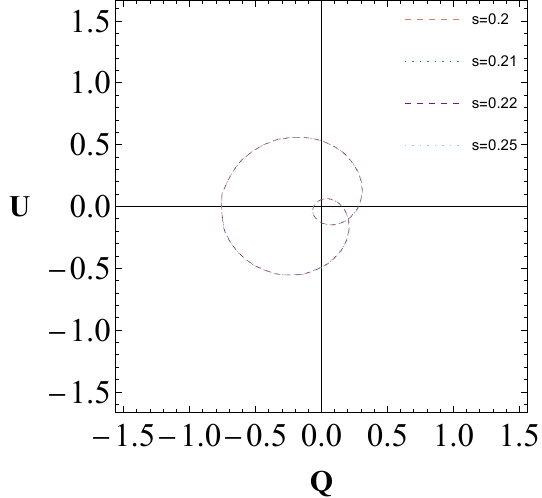}}
\subfigure[$\theta=40^{\circ}$]{\includegraphics[scale=0.475]{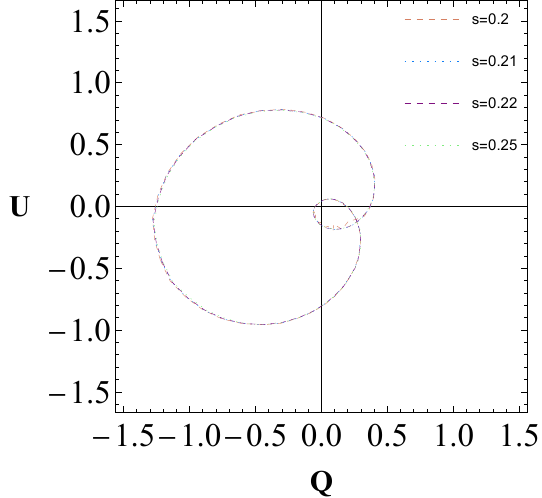}}
\subfigure[$B=(0,0.87,0.1,0)$]{\includegraphics[scale=0.475]{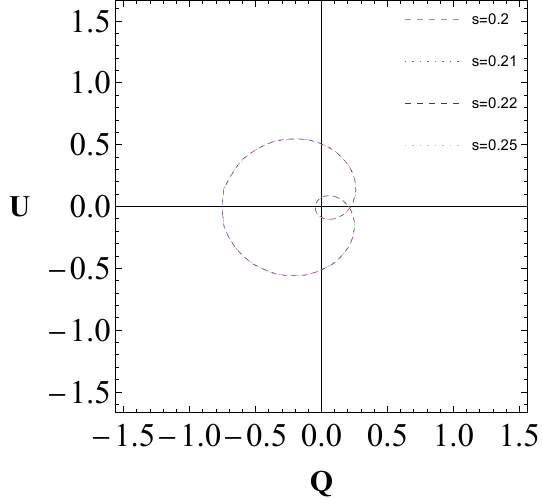}}
\subfigure[$B=(0,0.87,0.3,0)$]{\includegraphics[scale=0.475]{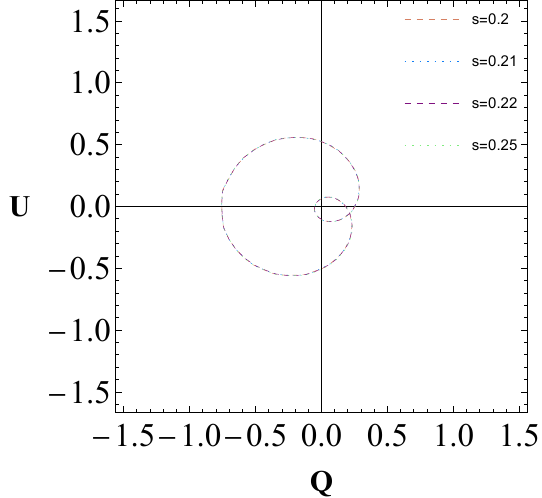}}
\subfigure[$B=(0,0.87,0.5,0)$]{\includegraphics[scale=0.475]{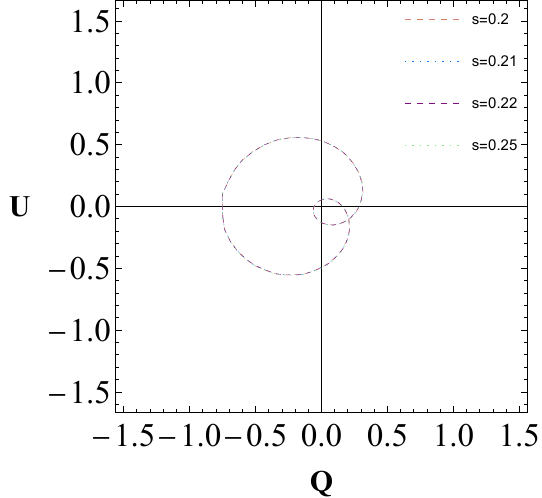}}
\subfigure[$B=(0,0.2,0.5,0)$]{\includegraphics[scale=0.475]{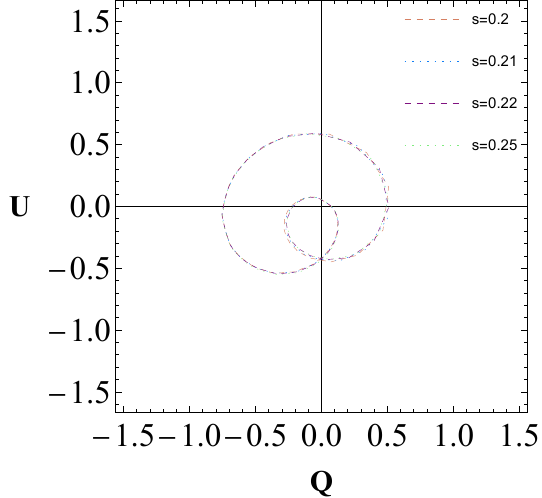}}
\subfigure[$B=(0,0.5,0.5,0)$]{\includegraphics[scale=0.475]{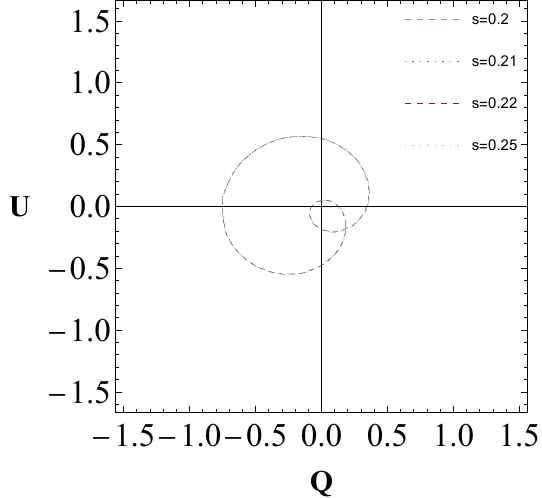}}
\subfigure[$B=(0,0.8,0.5,0)$]{\includegraphics[scale=0.475]{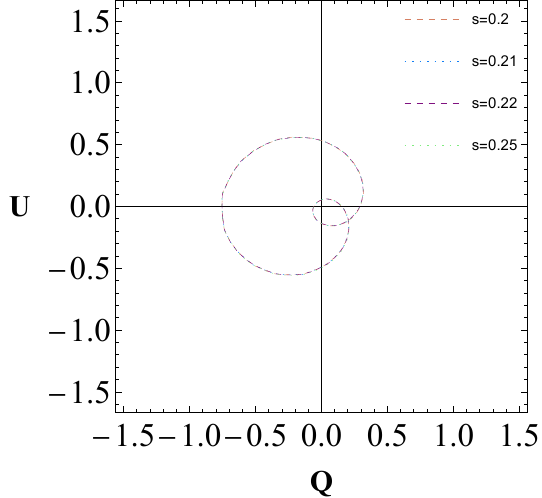}}
\caption{\label{fig4} The $Q-U$ planes for boson stars \( S_s\mathrm{BS1} \) to \( S_s\mathrm{BS4} \). The top row varies \(\theta = 20^\circ\), \(30^\circ\), \(40^\circ\) (left to right) with fixed \({B} = (0, 0.87, 0.5, 0)\). The middle row varies \( B_\theta = 0.1\), \(0.3\), \(0.5\) (left to right) at fixed \(\theta = 30^\circ\). The bottom row varies \( B_r = 0.2\), \(0.5\), \(0.8\) (left to right) at fixed \(\theta = 30^\circ\). }
\end{figure}

These trends demonstrate that variations in the $s$ parameter fundamentally alter the magneto-optical response of boson stars. Higher $s$ values amplify sensitivity to both geometric ($\theta$) and magnetic ($B_\theta$, $B_r$) parameters, particularly enhancing radial-field-induced depolarization. The resulting $s$-dependent polarization signatures introduce an additional diagnostic dimension for distinguishing horizonless compact objects from black holes, complementing the previously established $\phi_0$-dependent indicators.

Figure 5 explores the influence of the alteration of magnetic charge $q$ on the polarization image of the boson star. Each row corresponds to a distinct boson star configuration labeled from $S_{q}\mathrm{BS}_{1}$ to $S_{q}\mathrm{BS}_{4}$, which is characterized by an increasing magnetic charge. Specifically, the magnetic charge values are $q = 0.8$ (top row), $q = 0.83$ (second row), $q = 0.87$ (third row), and $q = 0.9$ (bottom row).
The columns display results for inclination angles $\theta = 1^\circ$ (leftmost), $\theta = 30^\circ$ (second), $\theta = 60^\circ$ (third), and $\theta = 80^\circ$ (rightmost). The electromagnetic sector parameter $s$ and the central scalar field amplitude $\phi_{0}$ are fixed across all panels at $s = 0.5$ and $\phi_{0} = 0.6$, respectively. The figure reveals systematic changes in the optical morphology of Bardeen boson stars as both magnetic charge $q$ and inclination angle $\theta$ vary. At near face-on orientations ($\theta = 1^\circ$), the accretion disk exhibits approximate rotational symmetry with concentric bright rings surrounding a central shadow. As the inclination increases ($\theta = 30^\circ$ and $60^\circ$), pronounced asymmetry develops, with relativistic Doppler boosting enhancing brightness along the approaching side of the disk. In edge-on configurations ($\theta = 80^\circ$), distinct vertical lensing arcs appear above and below the equatorial plane. Progressive enhancement of the magnetic charge from $q = 0.8$ to $q = 0.9$ expands the central shadow region while simultaneously enlarging the radius of the first bright photon ring. Higher $q$ values strengthen gravitational lensing effects, particularly evident at $\theta = 60^\circ$ and $80^\circ$, where multiple mirrored disk images become increasingly prominent. Competing brightness variations are observed: intensified emission from the inner disk contrasts with diminished secondary lensing features at high inclinations. The vertical separation between lensing arcs in edge-on views increases consistently across the magnetic-charge parameter space, indicating stronger light-bending effects at larger $q$. These transformations, obtained with fixed $\phi_0 = 0.6$ and $s = 0.5$, confirm the distinct role of the magnetic charge in modifying both the spacetime geometry and the resulting emission morphology.

\begin{figure}[!h]
\centering 
\subfigure[$q=0.8,\theta=1^{\circ}$]{\includegraphics[scale=0.375]{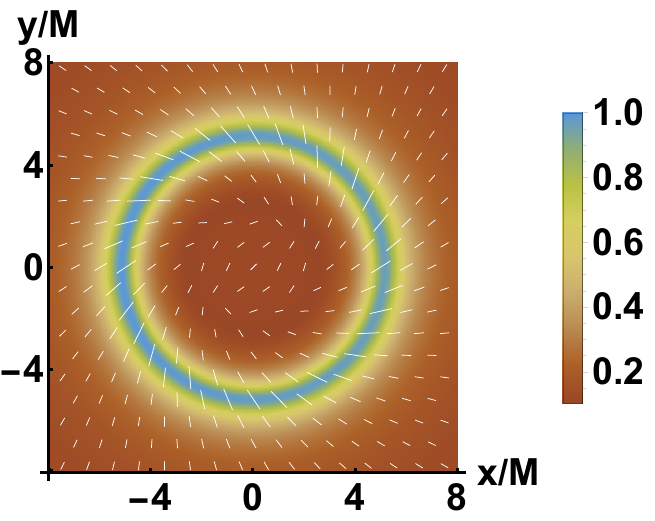}}
\subfigure[$q=0.8,\theta=30^{\circ}$]{\includegraphics[scale=0.375]{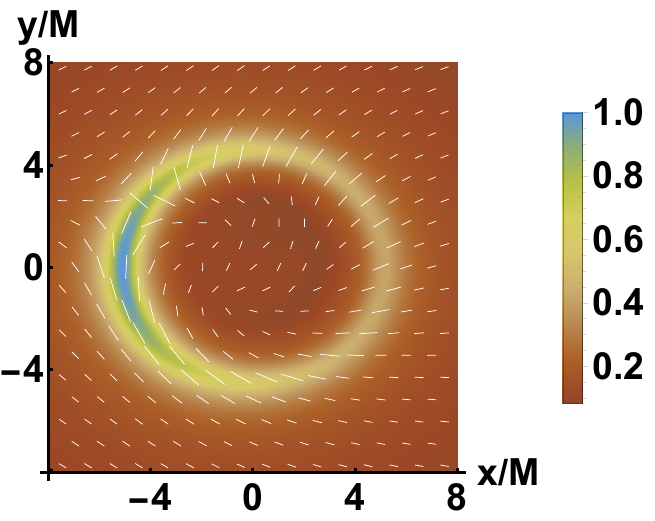}}
\subfigure[$q=0.8,\theta=60^{\circ}$]{\includegraphics[scale=0.375]{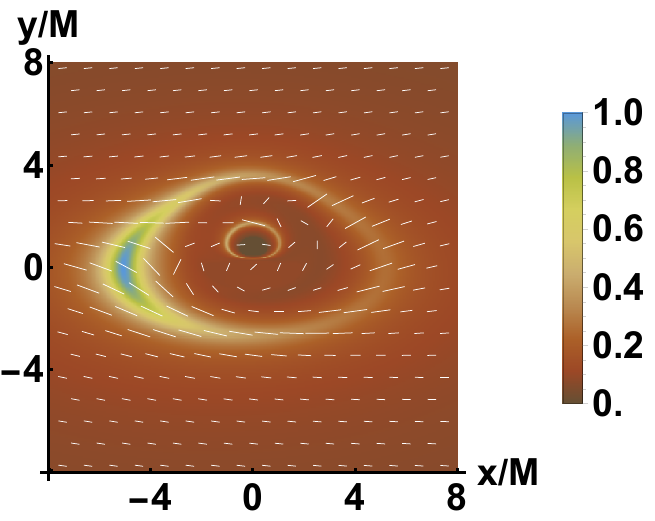}}
\subfigure[$q=0.8,\theta=80^{\circ}$]{\includegraphics[scale=0.375]{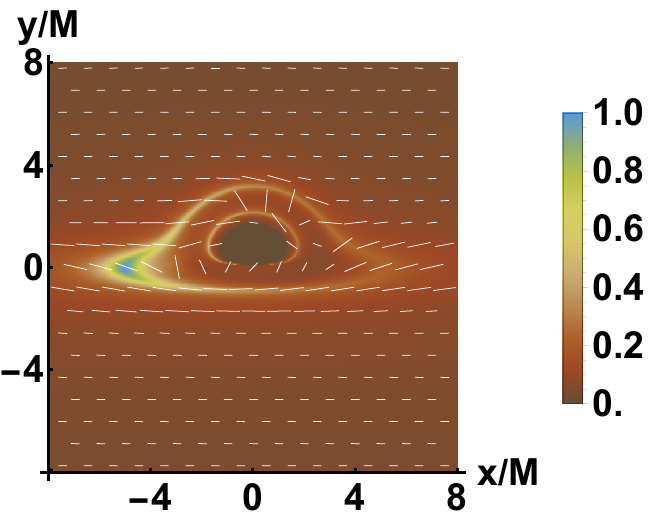}}
\subfigure[$q=0.83,\theta=1^{\circ}$]{\includegraphics[scale=0.375]{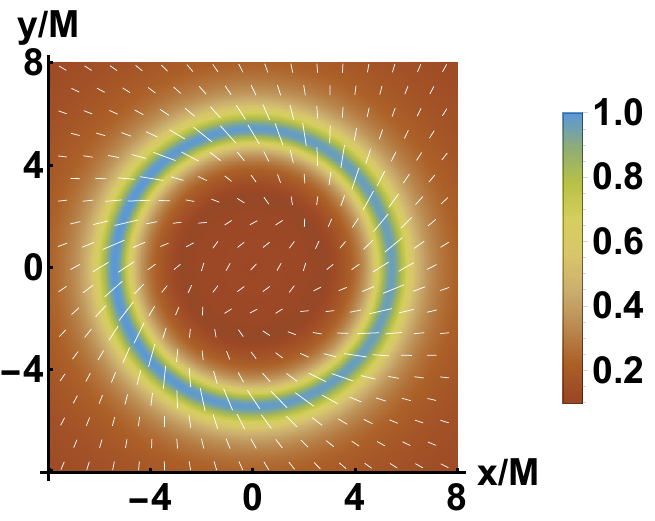}}
\subfigure[$q=0.83,\theta=30^{\circ}$]{\includegraphics[scale=0.375]{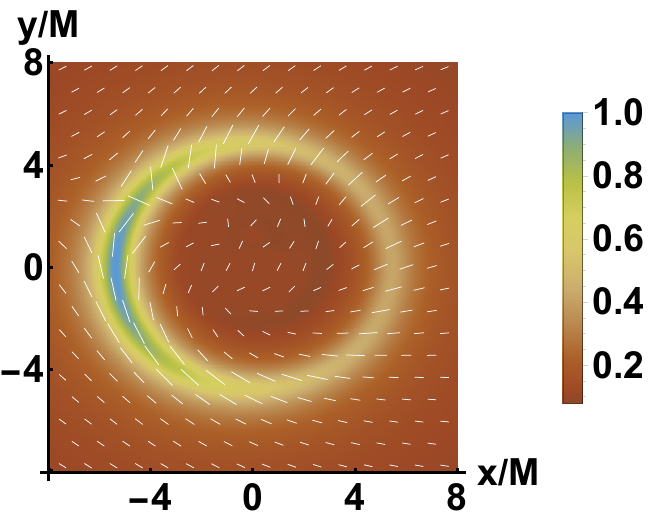}}
\subfigure[$q=0.83,\theta=60^{\circ}$]{\includegraphics[scale=0.375]{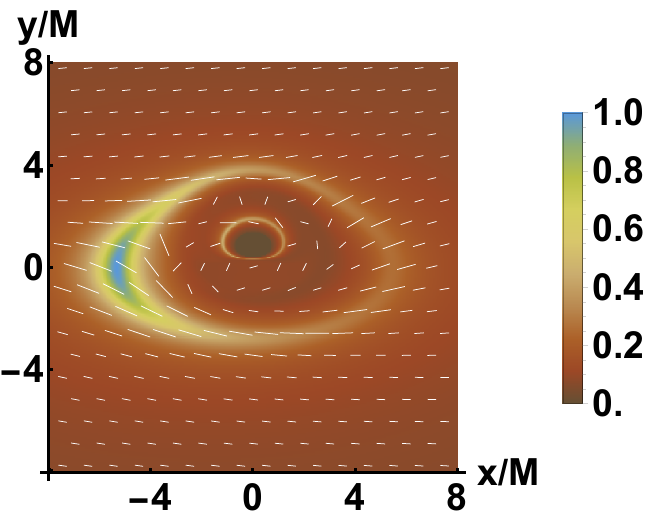}}
\subfigure[$q=0.83,\theta=80^{\circ}$]{\includegraphics[scale=0.375]{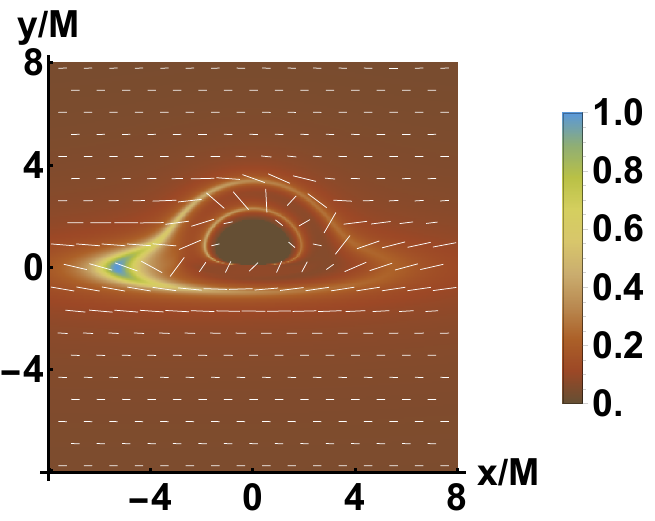}}
\subfigure[$q=0.87,\theta=1^{\circ}$]{\includegraphics[scale=0.375]{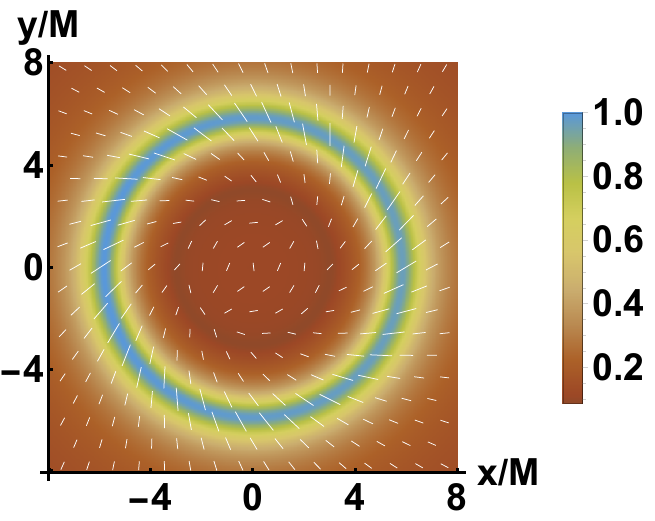}}
\subfigure[$q=0.87,\theta=30^{\circ}$]{\includegraphics[scale=0.375]{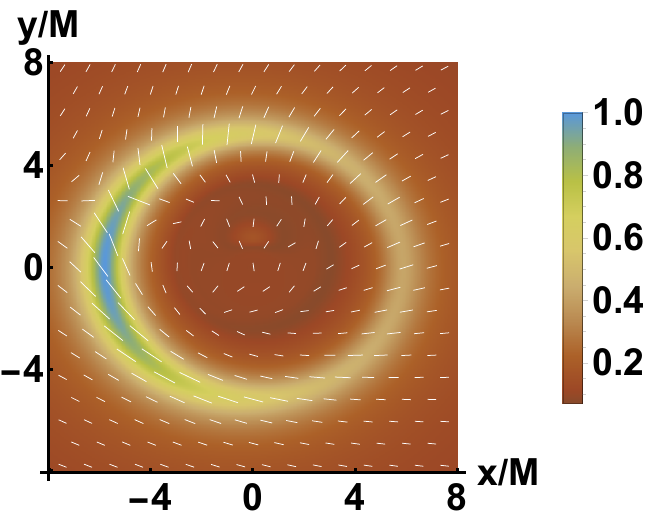}}
\subfigure[$q=0.87,\theta=60^{\circ}$]{\includegraphics[scale=0.375]{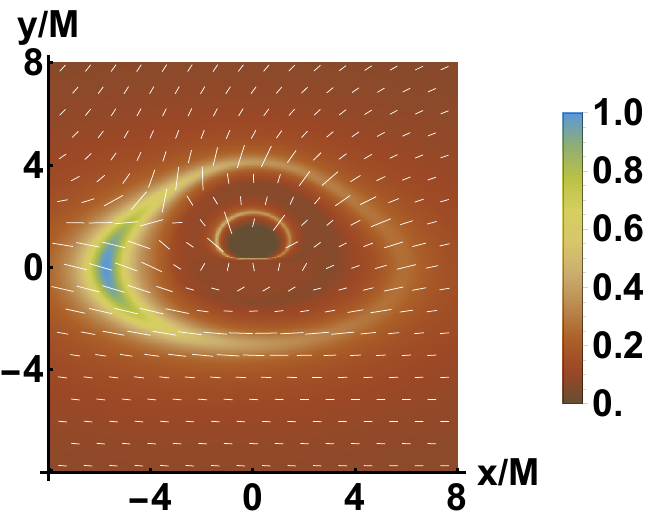}}
\subfigure[$q=0.87,\theta=80^{\circ}$]{\includegraphics[scale=0.375]{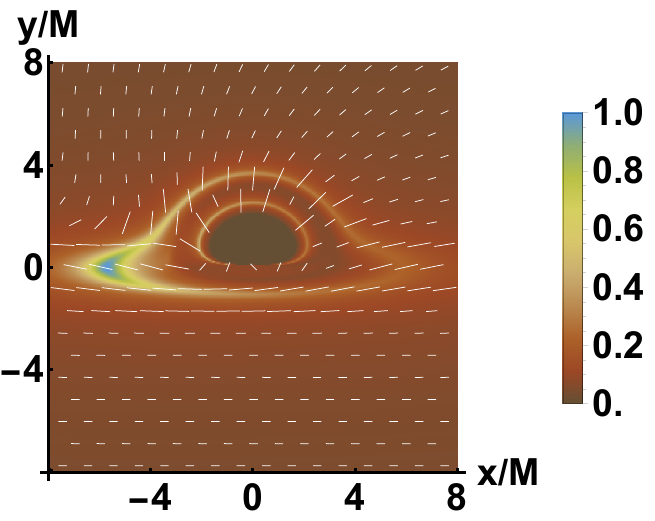}}
\subfigure[$q=0.9,\theta=1^{\circ}$]{\includegraphics[scale=0.375]{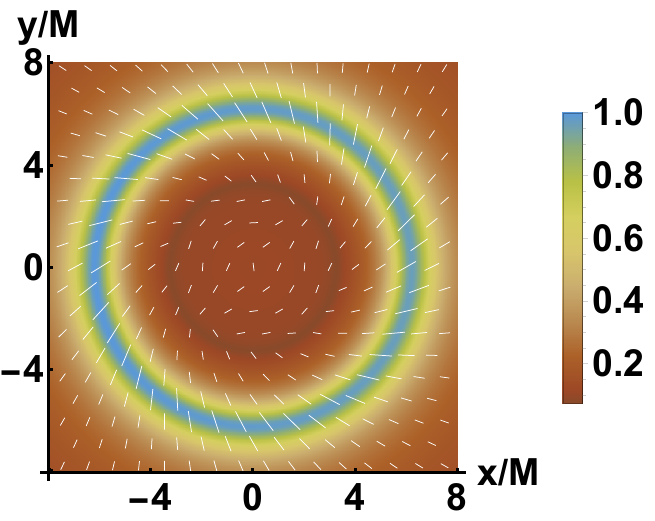}}
\subfigure[$q=0.9,\theta=30^{\circ}$]{\includegraphics[scale=0.375]{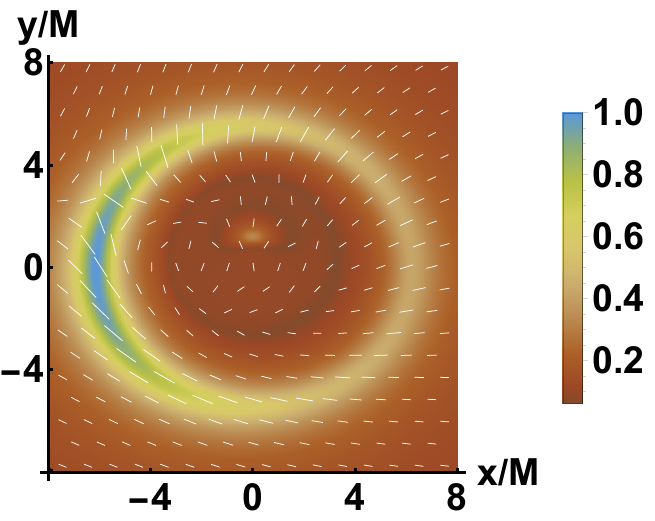}}
\subfigure[$q=0.9,\theta=60^{\circ}$]{\includegraphics[scale=0.375]{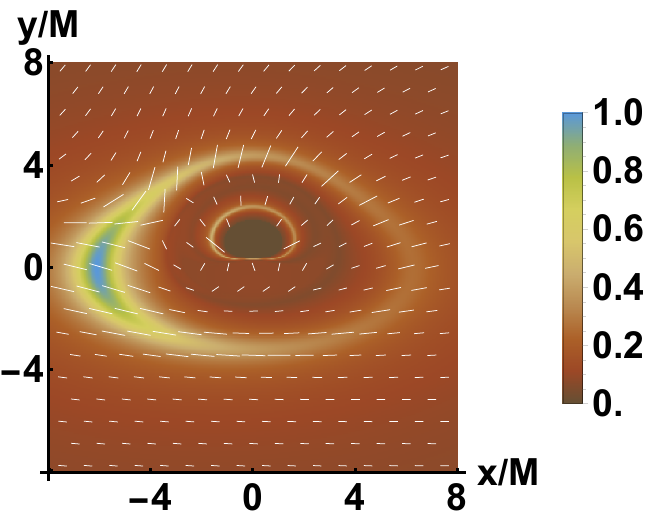}}
\subfigure[$q=0.9,\theta=80^{\circ}$]{\includegraphics[scale=0.375]{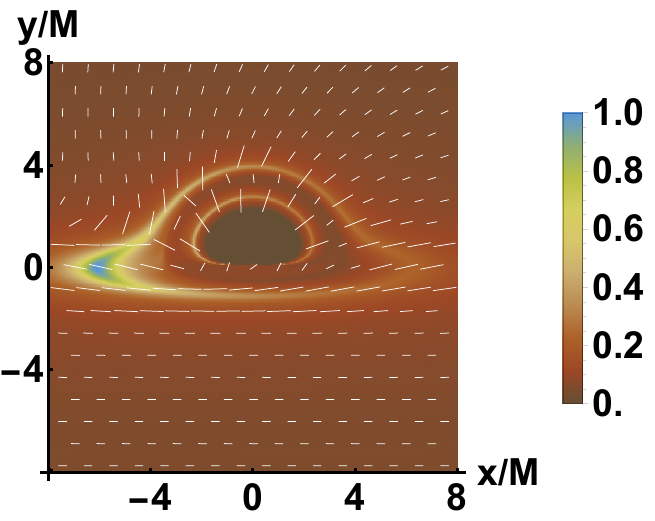}}
\caption{\label{fig5} The optical appearance (non-polarimetric) of boson stars under a thin accretion disk model with field of view \(\gamma_{\mathrm{fov}} = 5^\circ\). Images are organized into four rows for boson stars \(S_q\mathrm{BS1}\), \(S_q\mathrm{BS2}\), \(S_q\mathrm{BS3}\), and \(S_q\mathrm{BS4}\) with magnetic charge values \(q = 0.8\), \(0.83\), \(0.87\), \(0.9\) (top to bottom). Each row shows \(\theta = 1^\circ\), \(30^\circ\), \(60^\circ\), \(80^\circ\) (left to right). }
\end{figure}

\begin{figure}[!h]
\centering 
\subfigure[$\theta=20^{\circ}$]{\includegraphics[scale=0.475]{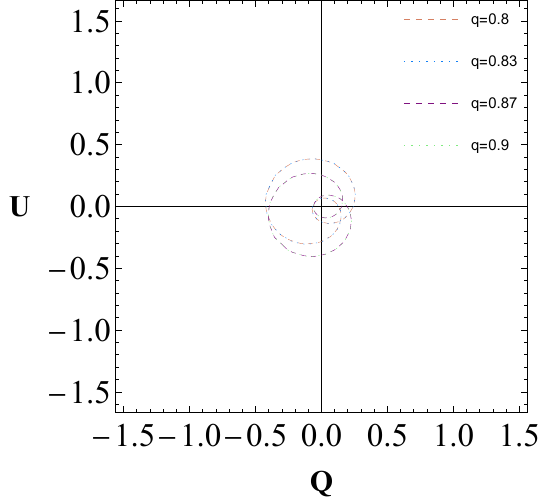}}
\subfigure[$\theta=30^{\circ}$]{\includegraphics[scale=0.475]{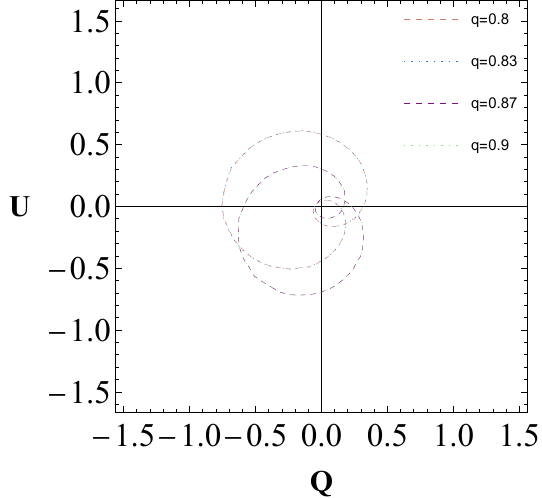}}
\subfigure[$\theta=40^{\circ}$]{\includegraphics[scale=0.475]{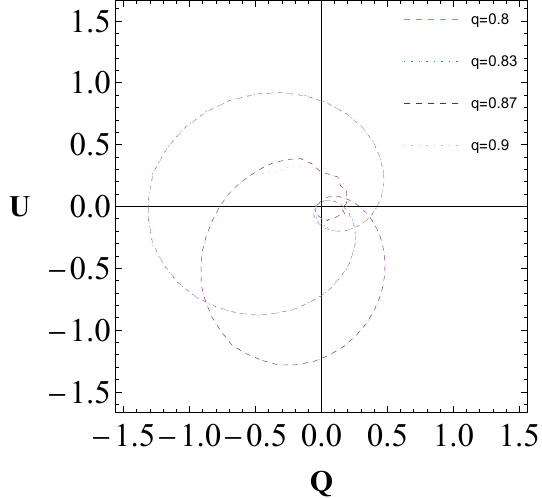}}
\subfigure[$B=(0,0.87,0.1,0)$]{\includegraphics[scale=0.475]{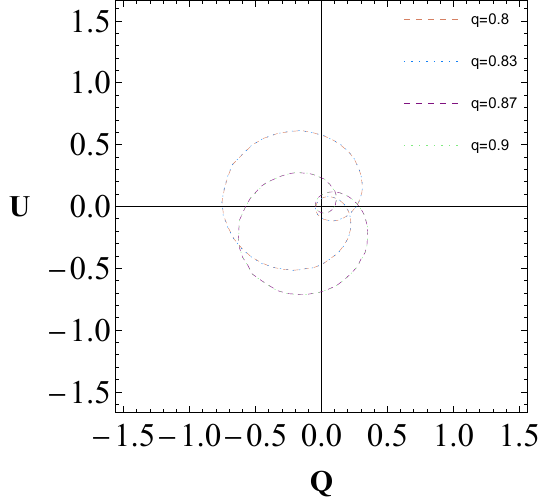}}
\subfigure[$B=(0,0.87,0.3,0)$]{\includegraphics[scale=0.475]{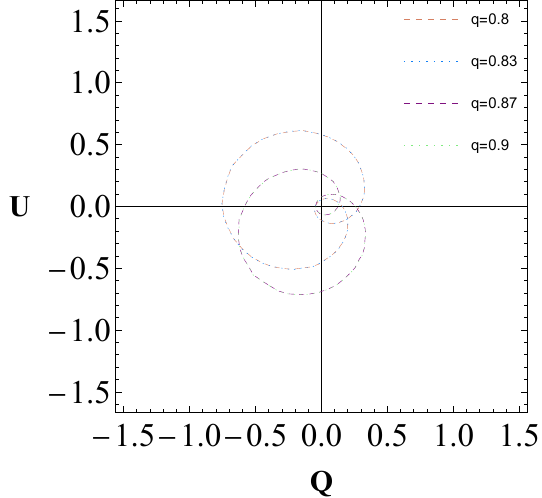}}
\subfigure[$B=(0,0.87,0.5,0)$]{\includegraphics[scale=0.475]{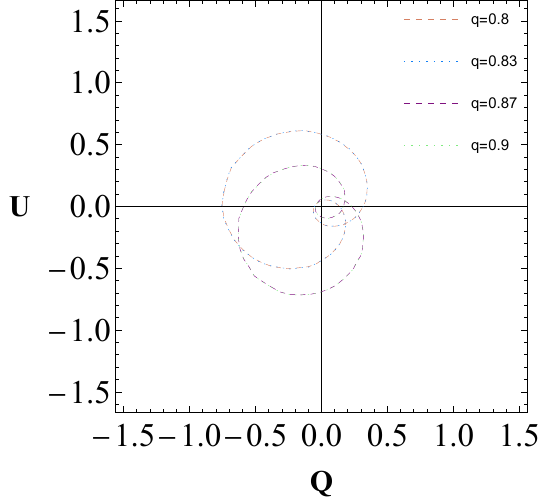}}
\subfigure[$B=(0,0.2,0.5,0)$]{\includegraphics[scale=0.475]{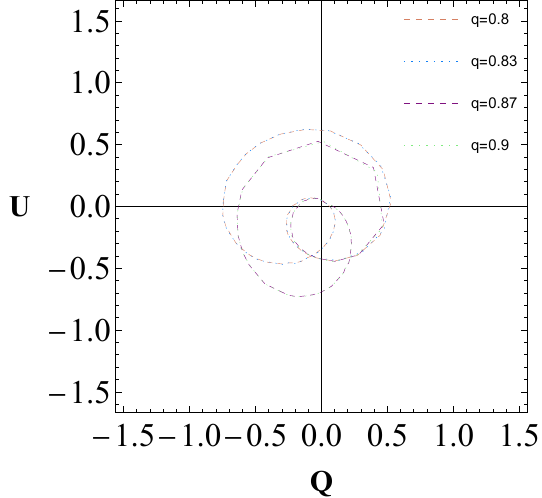}}
\subfigure[$B=(0,0.5,0.5,0)$]{\includegraphics[scale=0.475]{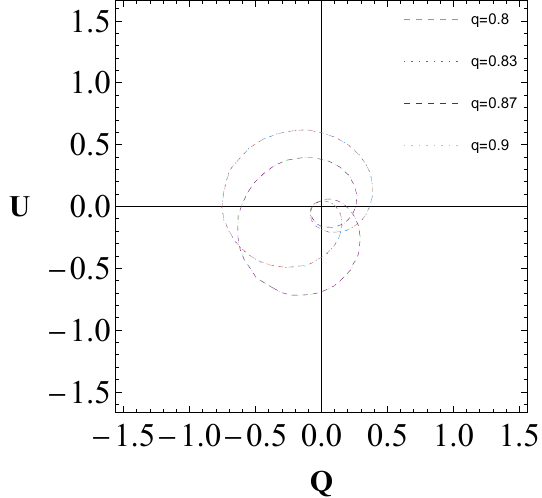}}
\subfigure[$B=(0,0.8,0.5,0)$]{\includegraphics[scale=0.475]{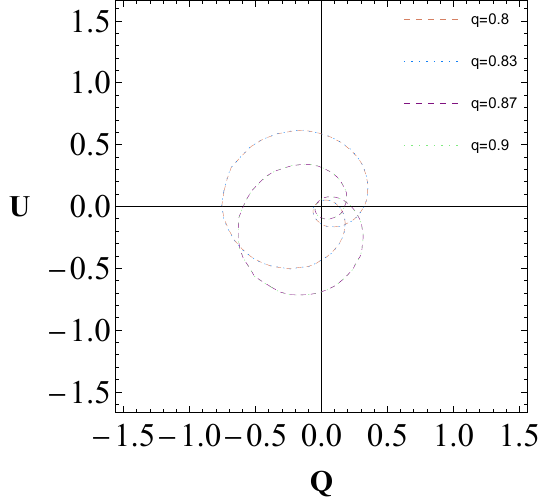}}
\caption{\label{fig6} The $Q-U$ planes for boson stars \(S_q\mathrm{BS1}\) to \(S_q\mathrm{BS4}\). The top row varies \(\theta = 20^\circ\), \(30^\circ\), \(40^\circ\) (left to right) with fixed \({B} = (0, 0.87, 0.5, 0)\). The middle row varies \(B_\theta = 0.1\), \(0.3\), \(0.5\) (left to right) at fixed \(\theta = 30^\circ\). The bottom row varies \(B_r = 0.2\), \(0.5\), \(0.8\) (left to right) at fixed \(\theta = 30^\circ\). }
\end{figure}

Figure~\ref{fig6} presents the polarization characteristics in the Stokes $Q-U$ plane for magnetically charged boson stars ($S_q\mathrm{BS1}$ through $S_q\mathrm{BS4}$) under varying parameters. The top row illustrates the dependence on inclination angle at a fixed magnetic field configuration $B = (0,\,0.87,\,0.5,\,0)$, showing systematic evolution of polarization loops as $\theta$ increases from $20^\circ$ to $40^\circ$. The middle row demonstrates the effect of the poloidal magnetic field component $B_\theta$ (ranging from 0.1 to 0.5) at fixed $\theta = 30^\circ$, where increasing $B_\theta$ progressively distorts the loop symmetry and enhances its eccentricity. The bottom row explores variations in the radial magnetic field component ($B_r = 0.2,\,0.5,\,0.8$) at $\theta = 30^\circ$, revealing characteristic loop rotations and centroid shifts toward the $Q$-axis with increasing $B_r$. Across all panels, higher magnetic-charge models ($S_q\mathrm{BS3}$, $S_q\mathrm{BS4}$) exhibit more compact polarization distributions and greater sensitivity to parameter variations than their lower-charge counterparts. Specifically, the inclination sequence (top row) shows a compact, nearly circular polarization loop centered near the origin at $\theta = 20^\circ$; increasing to $\theta = 30^\circ$ elongates the loop along the negative $Q$-axis while introducing mild asymmetries; at $\theta = 40^\circ$, the loop transforms into an open arc structure with noticeable dispersion along the $U$-axis. In the middle row, variations in $B_\theta$ reveal the progressive breakdown of loop symmetry: for $B_\theta = 0.1$, the loop remains symmetric but offset toward positive $Q$; at $B_\theta = 0.3$, pronounced distortion and multiple crossing points appear with increasing eccentricity; and at $B_\theta = 0.5$, the pattern fragments into two distinct clusters separated along the $U$-axis. The bottom row further demonstrates how the radial component modifies polarization morphology: $B_r = 0.2$ yields an elliptical loop rotated approximately $30^\circ$ counterclockwise; $B_r = 0.5$ shifts the centroid toward negative $Q$ while flattening the distribution; and $B_r = 0.8$ generates a figure-eight structure exhibiting clockwise rotation. Throughout all cases, higher-$q$ configurations ($S_q\mathrm{BS3}$, $S_q\mathrm{BS4}$) consistently display tighter polarization clustering and more pronounced parameter sensitivity than low-$q$ models, confirming the strong coupling between magnetic charge and polarization topology in horizonless boson stars.

\begin{figure}[!h]
	\centering 
	
	\subfigure[$\theta=30^{\circ}$]{\includegraphics[scale=0.37]{Q0902.pdf}}
	\subfigure[$\theta=80^{\circ}$]{\includegraphics[scale=0.37]{Q0904.pdf}}\\[0.5em]
	\subfigure[$\theta=30^{\circ}$]{\includegraphics[scale=0.37]{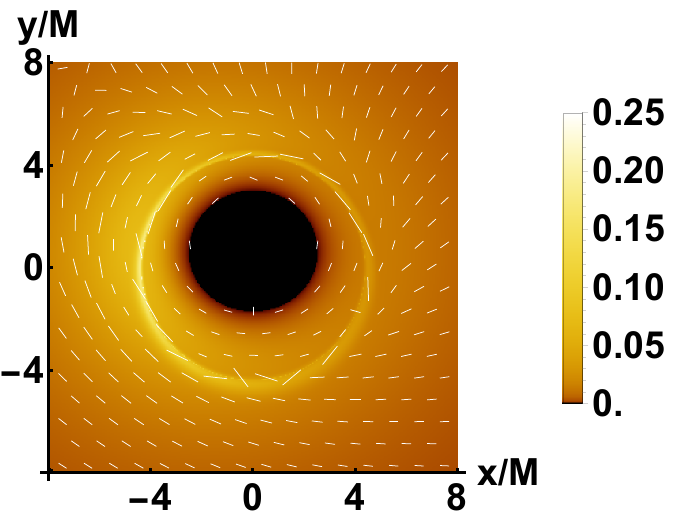}}
	\subfigure[$\theta=80^{\circ}$]{\includegraphics[scale=0.37]{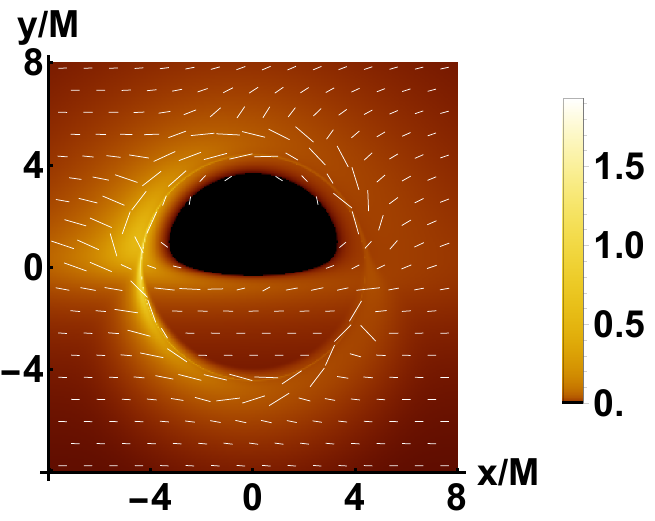}}
	\caption{\label{fig7}Comparison of the optical images of the boson star with the shadow of a Schwarzschild black hole. Panels (a) and (b) correspond to $S_q\mathrm{BS4}$, and panels (c) and (d) correspond to the Schwarzschild black hole.}
\end{figure}

Finally, we compare the optical images of the boson star with the shadow of a Schwarzschild black hole, taking $S_q\mathrm{BS4}$ as an example. Figure~\ref{fig7} shows the numerical results. As the observer inclination increases from $30^\circ$ to $80^\circ$, the images of both compact objects exhibit progressively more pronounced asymmetries. In the optical images, the Schwarzschild black hole displays a bright ring structure corresponding to the photon ring, which is absent in the boson star spacetime. The bright ring in the boson star images is mainly formed by the direct image. Furthermore, a central dark region also appears in the boson star images. However, it is significantly brighter than the shadow region of the Schwarzschild black hole, because the black hole has an event horizon while the boson star does not, allowing radiation to pass through the interior region.

\section{Conclusions and discussions}
\label{conclusion}
In this work, we have systematically investigated the polarized imaging properties of non-topological soliton Bardeen boson stars within a thin accretion disk framework by solving the coupled Einstein nonlinear electrodynamics complex scalar field equations. Our analysis demonstrates that the polarization patterns of these horizonless compact objects are highly sensitive to several key parameters, including the initial scalar field amplitude $\phi_0$, magnetic charge $q$, observer inclination angle $\theta$, and the magnetic-field geometry particularly the relative strengths of its radial ($B_r$) and poloidal ($B_\theta$) components.

We demonstrate that increasing the observer inclination angle produces asymmetric polarization structures with enhanced intensity along the projected equatorial plane, whereas stronger scalar fields generate expanded polarization regions accompanied by reduced overall intensity, a behavior reminiscent of quantum-corrected black holes. Distinctively, the electric vector position angle (EVPA) exhibits spiral-like morphologies at intermediate inclinations, showing a possible characteristic feature of boson stars.

Crucially, the Stokes $Q-U$ plane analysis reveals fundamental distinctions between boson stars and black holes. Boson stars exhibit loop fragmentation and depolarization under enhancements of the radial magnetic field, particularly at high scalar amplitudes, a phenomenon absent in the black hole case considered here, where the event horizon removes contributions from the interior emitting region. In contrast, vertical (poloidal) magnetic fields produce loop contraction and centroid migration without fragmentation. These polarization behaviors may serve as potential indicators for distinguishing horizonless compact objects from black holes.
Furthermore, variations in the nonlinear electrodynamics parameter $s$ and the magnetic charge $q$ significantly influence the polarization response, enhancing its sensitivity to both geometric and magnetic-field configurations.
This study provides a qualitative basis for further studies of polarized emission from horizonless compact objects in strong gravitational fields.

\vspace{10pt}

\noindent {\bf Acknowledgments}

\noindent
This work is supported by the National Natural Science Foundation of China (Grants Nos. 12505059, 12375043, 12575069), and the China Postdoctoral Science Foundation (Grants No.2025MD784184), and Chongqing Normal University Fund Project (Grants No. 26XLB001).

\end{document}